\documentclass[twocolumn,english,aps,pra,superscriptaddress,letterpaper]{revtex4-1}
\usepackage{amsmath,amssymb,graphicx}
\usepackage{textcomp} 
\usepackage{amsmath}
\usepackage[caption=false]{subfig}
\renewcommand{\vec}[1]{\boldsymbol{#1}} 
\usepackage{color}
\usepackage{hyperref}
\usepackage{cleveref}

\begin{document}
\title{Brillouin Optomechanics in Coupled Silicon Microcavities}

\author{Yovanny A. V. Espinel}
\author{Felipe G. Santos}
\author{Gustavo O. Luiz}
\author{Thiago P. M. Alegre}\email{alegre@ifi.unicamp.br}
\author{Gustavo S. Wiederhecker}\email{gustavo@ifi.unicamp.br}
\affiliation{Gleb Wataghin Physics Institute, University of Campinas, 13083-859 Campinas, SP, Brazil}

\begin{abstract}
The simultaneous control of optical and mechanical waves has enabled a range of fundamental and technological breakthroughs, from the demonstration of ultra-stable frequency reference devices to the exploration of the quantum-classical boundaries in laser-cooling experiments. More recently, such an opto-mechanical interaction has been observed in integrated nano-waveguides and microcavities in the Brillouin regime, where short-wavelength mechanical modes scatters light at several GHz. Here we engineer coupled optical microcavities spectra to enable a low threshold  excitation of mechanical travelling-wave modes through backward stimulated Brillouin scattering. Exploring the backward scattering we propose microcavity designs supporting super high frequency modes ($\sim25$~GHz) an large optomechanical coupling rates ($g_0/2\pi \sim 50$~kHz).
\end{abstract}
\flushbottom
\maketitle 
\newcommand{\nocontentsline}[3]{}
\newcommand{\tocless}[2]{\bgroup\let\addcontentsline=\nocontentsline#1{#2}\egroup}

\newcommand{\Gij}[3]{\frac{\left<\vec{\mathcal{E}}_{#1}|\delta\tilde #2|\vec{\mathcal{E}}_{#3}\right>}{\left<\vec{\mathcal{E}}_{#1}|\epsilon|\vec{\mathcal{E}}_{#3}\right>}}
\newcommand{\appropto}{\mathrel{\vcenter{
  \offinterlineskip\halign{\hfil$##$\cr
    \propto\cr\noalign{\kern2pt}\sim\cr\noalign{\kern-2pt}}}}}
\tocless{\section*{Introduction}}
Brillouin scattering occurs due to the interaction of optical and mechanical waves and it leads to the inelastic scattering of pump photons to Doppler red-shifted (Stokes) or blue-shifted (anti-Stokes) photons. In optical waveguides and microcavities this interaction occurs due to a combination of the photo-elastic effect~\cite{Boyd:2050257}, induced by strain, and moving-boundary effect caused by the mechanical mode distortion of the optical boundaries~\cite{Johnson:2002tp}. These two scattering processes are strongly influenced by optical and mechanical properties of the confining structure and can be tailored for various applications. For instance, the generation of anti-Stokes photons, which is accompanied by destruction of phonon quanta, can be used to cool down mechanical modes in optical cavities~\cite{Chan:2011dy,Anonymous:ARK9m6Hv}; whereas the generation of stokes photons, which create phonons (heating), may foster the development of high-coherence lasers, ultra-stable radio frequency (RF) synthesizers~\cite{Li:2012bfa,Smith:1991co,Debut:2000jz,Gross:2010jg}, and broadband tuning of RF filters~\cite{Marpaung:2014vf}. Such confinement-enhanced optomechanical interaction has been observed as Stimulated Raman-like~\cite{Dainese:2006ta} and Brillouin scattering in a range of photonic structures~\cite{Dainese:2006tj,Wiederhecker:2008tu,Kang:2009dja,Pant:2011ih,Rakich:2012et,Shin:2013fr,VanLaer:2015jk,Wolff:2015ji} \textemdash~where  both energy and momentum conservation are directly fulfilled.
  In microcavities, however, the short roundtrip length and narrow linewidth further constrain the conservation laws, requiring both pump and scattered waves to be resonant with the optical cavity modes in order to ensure efficient Brillouin scattering. These constraints have limited the current cavity demonstrations of Brillouin scattering either to mm-scale cavities~\cite{Li:2012bfa,Lee:2012hn,Grudinin:2009io,Lin:2014kc}, whose optical free-spectral range matches the mechanical resonant frequency; or heavily multimode micro-cavities, whose distinct transverse optical modes~\cite{Bahl:2012hf,Bahl:2013eb,Bahl:2012jm,Tomes:2009iy} frequency difference accidentally matches the mechanical frequency, both at the cost of reduced optomechanical coupling. 

\begin{figure}[!ht]
\centering
\includegraphics[scale=1.0]{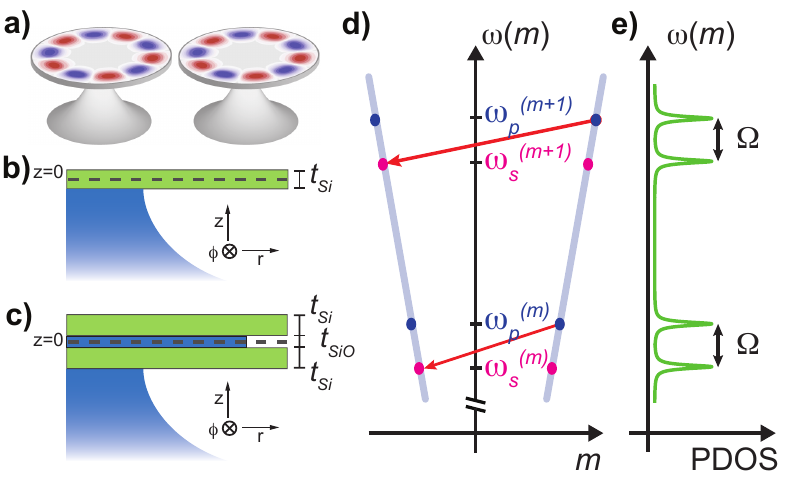}
\caption{\small{Backward Brillouin scattering in a compound microcavity system. \textbf{a)} Schematic of the compound microcavity system based on microdisk cavities. The colorscale represents one optical coupled mode of the structure. Single-disk \textbf{b)} and double-disk \textbf{c)} cavity designs; $t_\text{Si} = 250$~nm, $t_\text{SiO} = 100$~nm, the Si and SiO radius corresponding to $5$ and $3.8$~\textmu m, respectively. \textbf{d)} Optical dispersion diagram schematic, the optical resonances are represented by discrete (red and blue) points lying along the bulk dispersion curves (solid lines). Each pair of red and blue resonances are frequency split due to evanescent interaction in the compound system; the superscript $m$ denotes the azimuthal order of each mode family. The arrows indicate possible  resonant optical transitions from the pump ($\omega_{\text{p}}$) to the Stokes mode ($\omega_{\text{s}}$) due to BBS \textbf{e)} Photonic density of states (PDOS) at the pump and scattered waves when the optical frequency splitting matches the mechanical mode frequency $\Omega$.}}
\label{figure1}
\vskip -0.175in
\end{figure}

Here we explore a compound microcavity system based on silicon microdisk cavities and demonstrate its potential to drastically enhances backward Brillouin scattering (BBS) at tens of GHz. The compound microcavity scheme is illustrated in \cref{figure1}a and can ensure a doubly-resonant condition for the pump and stokes wave, yet preserving the small footprint necessary to achieve large optomechanical coupling and Brillouin gain. By engineering  the mechanical modes of single-disk ($sd$, \cref{figure1}b) and double-disk ($dd$, \cref{figure1}c)  optical microcavities to avoid cancellation between the photo-elastic (\textit{pe}) and moving-boundary (\textit{mb}) effect~ \cite{Florez:2016aa}, we demonstrate  that both cavity designs  could be exploited in the compound cavity scheme, offering a promising route towards the demonstration of low threshold backward stimulated Brillouin lasing in a CMOS-compatible platform. 

In backward Brillouin scattering the optical pump and the scattered Stokes waves propagate in opposite directions, resulting in a large wavevector mismatch that favors the interaction between light and short-wavelength propagating mechanical modes~\cite{Boyd:2050257}. In disk microcavities the optical and mechanical modes are azimuthally traveling waves with azimuthal dependence $\exp(\pm i m \phi)$ (here $m$ is an integer and $\phi$ the azimuthal angle). Therefore, a pump laser exciting an optical cavity mode with frequency and azimuthal number ($\omega_{\text{p}},m_{\text{p}}$) may be scattered into another optical mode ($\omega_{\text{s}},m_{\text{s}}$) through the interaction with a mechanical mode ($\Omega,M$), provided that both energy and momentum (phase-matching) are conserved, i.e  $\omega_{\text{s}}(m_{\text{s}})=\omega_{\text{p}}(m_{\text{p}}) \pm \Omega(M)$ and $m_{\text{s}}=m_{\text{p}}\pm M$.
 While in forward Brillouin scattering the phase-matching condition favors mechanical modes close to their cut-off condition $M=0$ ($m_{\text{s}}=m_{\text{p}}$), in backward Brillouin scattering (BBS) the scattered light frequency shift is proportional to the optical wavevector mismatch and can easily reach tens of GHz in solids, $\Omega \approx (M/r)V_\text{m}=(2m_\text{p}/r)V_\text{m}$, where $r$ is a typical cavity radius and $V_\text{m}$ is the mechanical mode phase velocity. In order to enhance BBS, such a  large frequency shift would require the pump wave to be detuned from the optical resonance by tens or even hundreds of linewidths \textemdash~in a single-resonance cavity such a large detuning would drop the benefits of the resonant cavity build-up for the pump wave.
 In the proposed compound cavity system, illustrated in \cref{figure1}a, the interaction between the optical modes (through their evanescent fields) leads to a frequency splitting that can be precisely controlled during microfabrication by adjusting the distance between the cavities. This scheme is illustrated in \cref{figure1}d with the pump wave tuned to the higher frequency coupled mode at $\omega_{\text{p}}$, while the lower frequency coupled mode is resonant with the scattered wave $\omega_{\text{s}}$, thus ensuring a high photonic density of states (PDOS) at the pump and scattered frequencies (see \cref{figure1}e).
\tocless{\section*{Results}}

\tocless{\subsection*{Brillouin interaction}}

The large azimuthal numbers involved in BBS imply that the phase-matched mechanical modes are localized near the cavity edge~\cite{Tamura:2009bx}, compared to low azimuthal number modes that are spread throughout the cavity, such an edge localization effectively increases the overlap between the optical and mechanical modes~\cite{Sturman:2015jy,Matsko:2005cz}. The mechanical mode induced strain and boundary deformation at the cavity edge Bragg scatter light and efficiently couple forward and backward propagating optical modes~\cite{Dostart:2015ju}. The resulting energy exchange between the optical pump and Stokes wave can be modeled using coupled mode theory~\cite{Matsko:2002vs,Agarwal:2013hl}, which leads to a set of coupled equations for the amplitudes of the pump wave, stokes wave, and mechanical wave (see Methods).
\begin{figure}[!ht]
\includegraphics[scale=1]{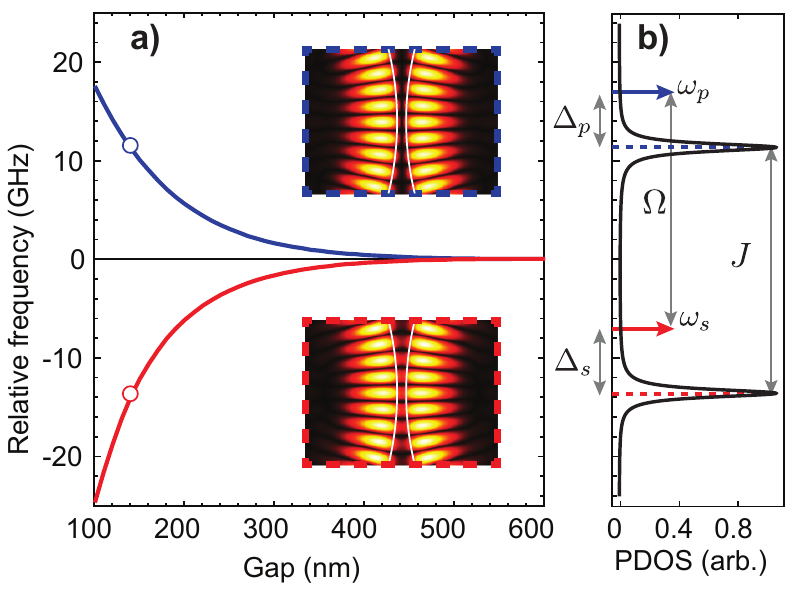}
\caption{\small{\textbf{Coupled cavity scheme.} \textbf{a)} Optical frequency splitting approximation to microdisk single cavity (\cref{figure1}b). Blue and red curves indicate the anti-symmetric and symmetric coupled modes, respectively. The insets show the calculated optical mode profiles for a 140 nm air-gap between the cavities (circle's labels). \textbf{b)} Photonic density of states obtained for the coupled cavity modes with a quality factor of $10^5$ and splitting rate $J=25$ GHz. The dashed blue and red lines indicate the optical mode resonant frequencies, $\omega_{{\text{p}}_0}$, $\omega_{{\text{s}}_0}$}, respectively.}
\label{figure2}
\vskip -0.175in
\end{figure}
We use the coupled mode theory to derive the threshold power necessary to achieve stimulated Brillouin lasing, which occurs when the Stokes photons gain induced by the optical pump suppresses the Stokes cavity mode loss. By assuming an undepleted pump the following expression can be derived for the threshold power~\cite{Matsko:2002vs} (see Supplementary Information),
\begin{equation}
	P_\text{th}=\frac{\hbar\omega_{\text{p}}\kappa_{\text{p}}^2}{4\,\mathcal{C} \kappa_{\text{e}}}\left[1+\left(\frac{\Delta_{\text{p}}}{\kappa_{\text{p}}/2}\right)^2\right]\left[1+\left(\frac{\Delta_{\text{s}}}{\kappa_{\text{s}}/2}\right)^2\right],
	\label{eq:threshold}
\end{equation}
where $\mathcal{C}=4(g_0^\text{c})^2/(\Gamma\kappa_{\text{s}})$ is the so-called single-photon cooperativity; $g_0^\text{c} $ is the vacuum optomechanical coupling rate for the compound cavity and $\Delta_{\text{s}}=\omega_{\text{s}}-\omega_{{\text{s}}_0}$ and $\Delta_{\text{p}}=\omega_{\text{p}}-\omega_{{\text{p}}_0}$ are the pump and stokes detuning (see \cref{figure2}); $\kappa_{\text{s}}$ and $\kappa_{\text{p}}$ are the corresponding total (intrinsic and extrinsic) loss rates, $\kappa_{\text{e}}$ is the extrinsic pump loss rate due to coupling to the driving mode of the bus waveguide.
When both pump and Stokes waves are resonant with coupled-cavity modes ($\Delta_{\text{s}},\Delta_{\text{p}}$)=0, the lowest threshold power is reached. Note that the Stokes photons are initially generated by spontaneous Brillouin scattering (due to thermally driven phonons) and therefore $\omega_{\text{s}}=\omega_{\text{p}}-\Omega$. 
When the pump and Stokes optical modes are separated by a frequency difference $J$, their detuning is given by  $\Delta_\text{s}=\Delta_{\text{p}}+(J-\Omega)$ (see \cref{figure2}b). Therefore,  the threshold power scales as $P_\text{th}\propto(1+4(J-\Omega)^2/\kappa_{\text{s}}^2)$   for a resonant pump ($\Delta_\text{p}=0$), and the minimum threshold occurs when the optical splitting precisely matches the mechanical frequency, i.e. $J=\Omega$. This threshold power scaling reveals the importance of the doubly-resonant condition ensured by the  compound cavity scheme. For instance, the minimum threshold power achievable using a standard single-resonance cavity occurs in the so-called sideband resolved limit ($\Omega_{\text{m}}\gg\kappa$) at an optimum pump detuning ($\Delta_\text{p}=\Omega$), this limit  can be obtained from \cref{eq:threshold} assuming a large cavity separation ($J\rightarrow 0$, then $\kappa_\text{s}\rightarrow \kappa_\text{p}$ and $g_{0}^{\text{c}}\rightarrow g_{0} $). Therefore a single-resonance cavity has a threshold power $(\Omega_{\text{m}}/\kappa_{\text{p}})^2$ larger than the proposed compound cavity doubly-resonant approach.
 For a typical $5$ \textmu m radius microdisk   $(\Omega_{\text{m}}/\kappa_{\text{p}})^2\approx 100$, a roughly  two-orders of magnitude higher threshold; where we assume an intrinsic quality factor of $2\times 10^5$ ($\kappa_\text{p,s}/2\pi\approx 960$~MHz) and Brillouin frequency $\Omega/2\pi=V_\text{l} M/r \approx 22$~GHz \textemdash~where $M/r\approx 4\pi (n_\text{eff}/\lambda)$ and $n_\text{eff}=1.73$ for the transverse-magnetic (TM) optical mode phase index ($\lambda=1.55$~\textmu m), $V_\text{l}$ is the Si bulk dilational wave velocity. 
	Such high mechanical frequencies at tens of GHz can also be readily matched to the optical resonance splittings accessible with either single or double-disk silicon cavities, in contrast  with larger  lower refractive index microcavities \cite{Grudinin:2009tm} whose frequency splitting lies in the MHz-range range.   For example, we show in \cref{figure2} the numerically calculated frequency splitting curves for a single disk (solid lines) silicon cavity (\cref{figure1}b, see Methods) that demonstrate frequency splitting at tens of GHz around 100 nm gap between the cavities.\\  

\tocless{\subsection*{Device design}}

\begin{figure*}[!ht]
\centering
\includegraphics{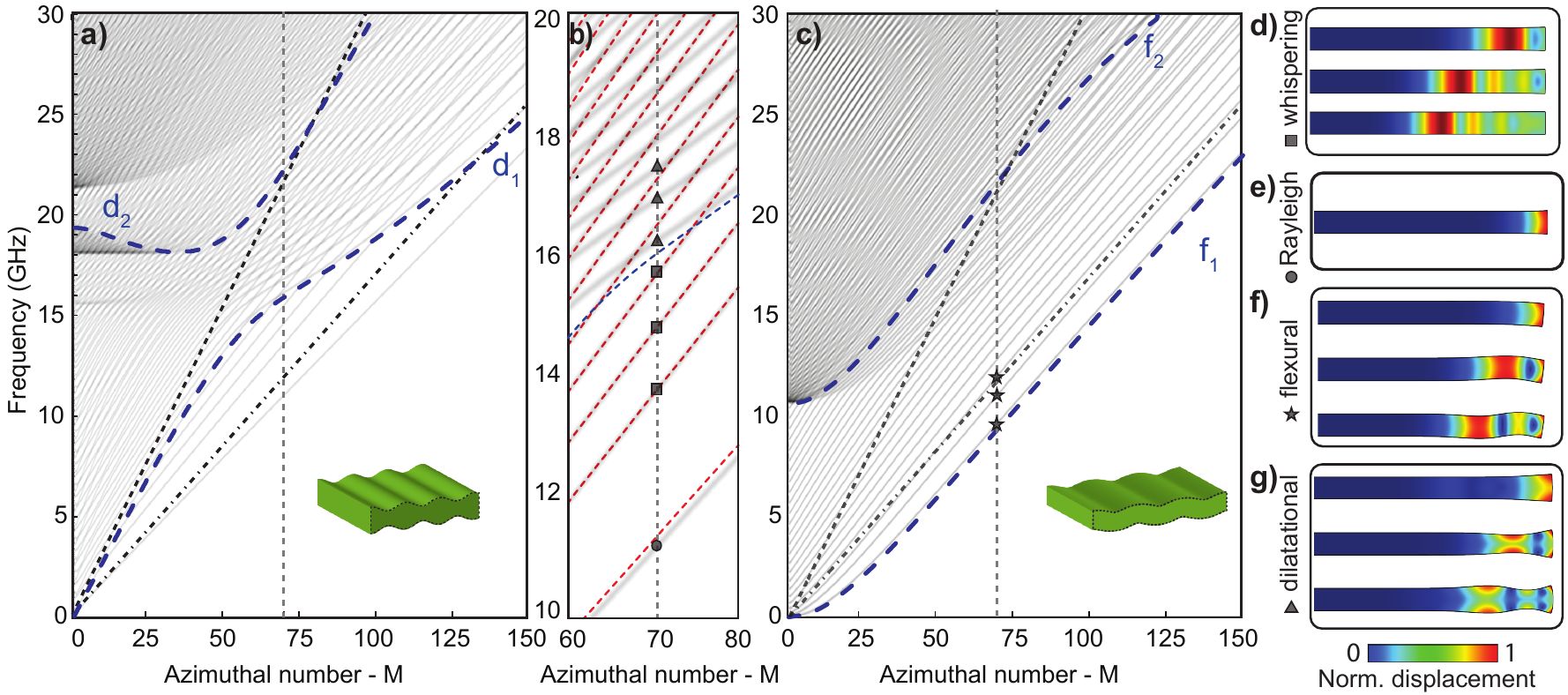}
\caption{\small{\textbf{Mechanical dispersion in the $sd$-cavity}. \textbf{a)} Dispersion diagram for even modes (grey curves). The dashed and dash-dotted black solid lines represent the longitudinal ($V_\text{l}=9660$ m/s) and transverse ($V_\text{t}=5340$ m/s) bulk Si acoustic velocities. The dashed blue lines represent the dispersion of the first two dilatational modes (d$_{1}$ and d$_{2}$) of a 250 nm thick silicon slab (inset shows to d$_{1}$ mode at $M=70$). The vertical dashed line ($M=70$) indicates the phase-matching azimuthal number for the TM optical mode in 1550 nm, $m_{\text{p}}=M/2=35$. \textbf{b)} Zoom for the dispersion of the even modes around of $M=70$. The red dashed lines represent the dispersion of the whispering gallery modes for an infinite cylinder with radius like the $sd$-cavity. The blue dashed line is the dispersion of the d$_{1}$ mode. The geometrical markers along to the vertical dashed line make reference to different families of the modes in the $sd$-cavity. \textbf{c)} Dispersion diagram for the odd modes (grey curves). The dashed blue lines represent the dispersion of the first two flexural modes (f$_{1}$ and f$_{2}$) of a 250 nm thick silicon slab, inset shows to f$_{1}$ mode at $M=70$ (The vertical dashed line). \textbf{d)} First modes of the whispering gallery family (square's markers in b)). \textbf{e)} Rayleigh mode (circle's marker in b)). \textbf{f)} First modes of the flexural family (star's markers in c)). \textbf{g)} First modes of the dilatational family (triangle's markers in b)).}}
\label{figure3}
\vskip -0.175in
\end{figure*}

We demonstrate the feasibility of our compound cavity scheme by investigating two designs that can achieve high optomechanical coupling rates and mechanical frequencies at tens of GHz, the $sd$ and $dd$ cavities shown in \cref{figure1}b,c. The mechanical dispersion is the starting point to infer general characteristics of the phase-matched mechanical modes that will lead to the Brillouin scattering in microdisk cavities.
Many aspects of the mechanical modes dispersion of $sd$  and $dd$ cavities  can be regarded as mixtures among whispering gallery modes of an infinite cylinder and Lamb-wave modes of a free-standing silicon slab~\cite{Tamura:2009bx,Dmitriev2014905}.
 The mechanical mode dispersion of a single disk for even and odd modes (with respect to $z=0$ plane \textemdash~see \cref{figure1}b) are shown in  \cref{figure3}a and \cref{figure3}c, respectively. The dispersion curves are calculated using an axisymmetric finite element method (see Methods).  In the $dd$-cavities, the mechanical modes are approximately symmetric/anti-symmetric combinations of the even and odd parity $sd$-cavity modes, therefore, the key characteristics of both designs may be inferred by inspecting only the $sd$ mode structure.

The mechanical modes of $sd$ cavities that may efficiently interact with optical modes can be divided in four groups: whispering gallery, Rayleigh, dilatational, and flexural modes. Their dispersion curves are signaled by markers in \cref{figure3}b,c while corresponding displacement profiles are shown in \cref{figure3}d-g. 
The whispering gallery  group ($w$-modes) modes are remarkably similar to modes of an infinite cylinder (not shown), as suggested by the excellent agreement between their dispersion curves (shown in \cref{figure3}b) and displacement profiles, which are essentially in the radial-azimuthal ($r\phi$) directions \textemdash~despite the very small thickness/radius ratio of our disk ($t/r=0.05$). The large shifting of the displacement peak radial position across the $w$-mode group, noticeable in \cref{figure3}d, already suggests a varying overlap with the optical mode. 
The Rayleigh ($r$-modes), dilational ($d$-modes), and flexural ($f$-modes) mode groups are signatures of the thin disk; their dominant radial-vertical ($rz$) displacement are noticeable in \cref{figure3}e-g.
The $r$-mode is a singleton group and has the lowest frequency dispersion branch and characterized by a phase velocity lower than both the longitudinal ($V_{\text{l}}$) and transverse ($V_{\text{t}}$) bulk velocities (\cref{figure3}a)~\cite{Tamura:2009bx,Dmitriev2014905}, as shown from the displacement profile in \cref{figure3}e; such a surface wave localization  compromises its overlap with the optical mode.
The slab-like nature of the $d$-modes is evidenced not only by  their  displacement profiles in \cref{figure3}g but also through their good agreement with the slab dilatational modes dispersion shown in \cref{figure3}b (blue-dashed curve). The onset of the distinct disk mechanical mode families in \cref{figure3}a is also well matched by slab-mode cutoff frequency. Based on their displacement profile, the $d$-mode group is likely to have modes with large overlap with optical modes.
On the other hand, the $f$-group resemble cantilever modes and is the only group with an odd symmetry relative to the $z=0$ plane, resulting in a negligible interaction with $sd$-cavity optical modes.      
In the $dd$-cavities however, the symmetric combination of upper and lower-disk $f$-modes strongly modulate the air-gap between the disks and also has the potential to strongly interact with the double-disk optical modes. These symmetric $f$-modes are similar to those explored in previous double-disk devices~\cite{Wiederhecker:2009ex,Zhang:2012us}, but due to their large azimuthal number they can readily vibrate in the 10 GHz frequency range.  
%

\begin{figure*}[!ht]
\centering
\includegraphics[scale=1]{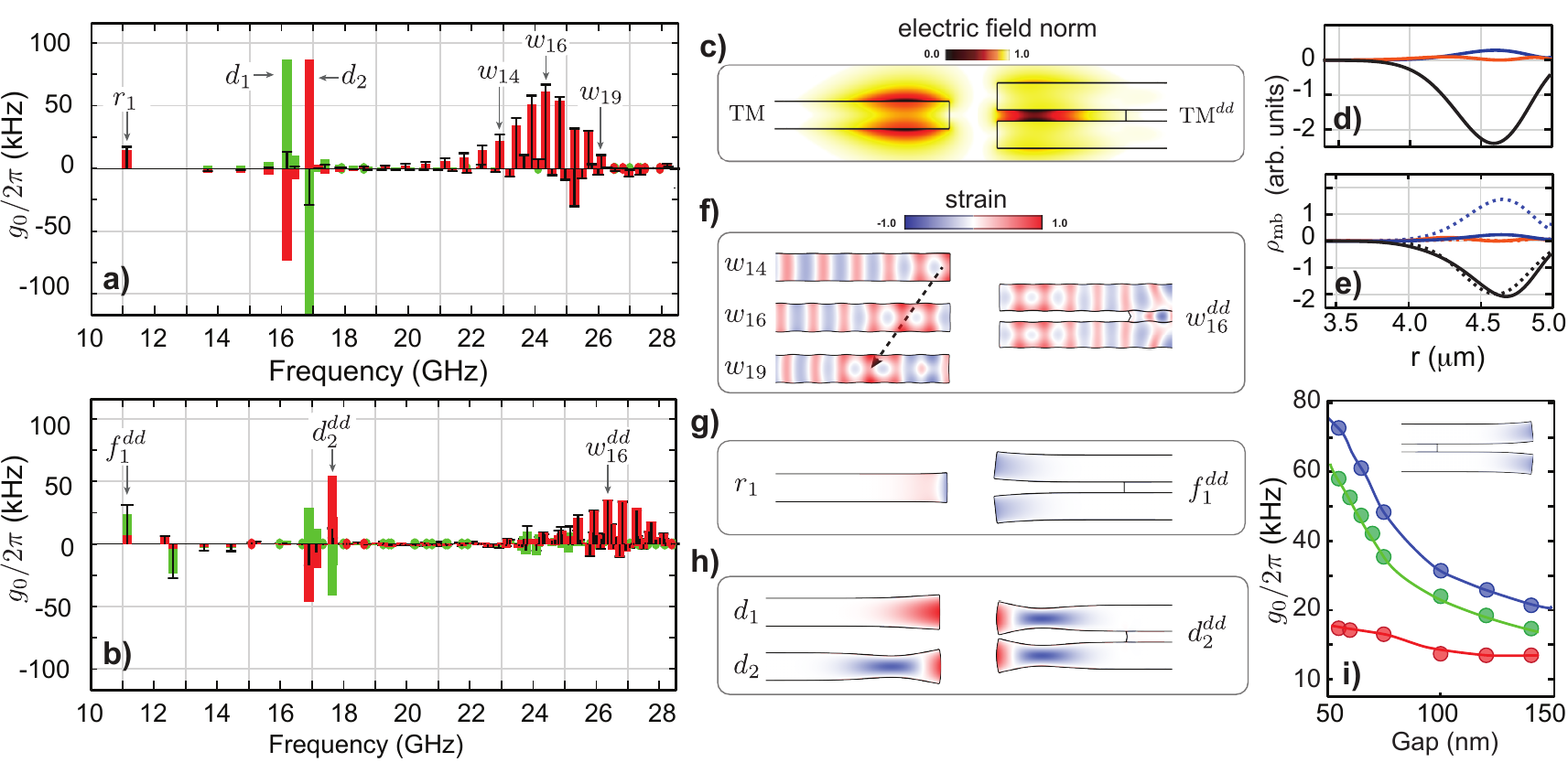}
\caption{\small{\textbf{a), b)} Optomechanical coupling rate (black bars) between the optical mode and the mechanical even modes (respect to z=0 plane in \cref{figure1}b-c) generated by the photoelastic $g_\text{pe}/2\pi$ (red) and moving boundary  $g_\text{mb}/2\pi$ (green) contributions in the $sd$ and $dd$ cavities, respectively. \textbf{c)} Electric field norm of the optical mode at 1550 nm for the \textit{sd}-cavity (TM) and the \textit{dd}-cavity (TM$^{dd}$). Contributions (solid lines) to optical weighting function $\rho_{\text{mb}}$ (\cref{eq:gmb_terms}) along the upper or lower $sd$-cavity boundaries \textbf{d)} and along the outer upper (solid lines) and slot-interior (dashed-lines) $dd$-cavity boundaries \textbf{e)}. The red, black and blue lines represent the contributions related essentially to the field components $E_{r}$, $E_\phi$ and $D_z$, respectively. \textbf{f)-h)} Dominant strain component (colorscale) and deformation (amplified) for the mechanical modes labeled in part a) and b). The strain component $S_{\phi\phi}$ is shown for the $w$-modes and $r$-modes ($r_{1}$), $S_{\phi z}$ for $f$-modes ($f_{1}^{dd}$) and $S_{zz}$ for the $d$-modes. \textbf{i)} Optomechanical coupling rate for $f_{1}^{dd}$ as a function of the double-disk gap. The blue, green and red markers represent the total, \textit{mb} and \textit{pe} optomechanical coupling rates, respectively. The blue, green and red solid lines only should be considered as a guide.}}
\label{figure4}
\vskip -0.175in
\end{figure*}

The spatial overlap between optical modes (\cref{figure4}c) and mechanical modes (\cref{figure3}d-f) is necessary but not sufficient for a large optomechanical coupling. The optomechanical interaction in these high refractive index structures occurs due to a combination of the photo-elastic effect~\cite{Boyd:2050257} and deformation of the cavity boundaries~\cite{Johnson:2002tp}.  
The calculated optomechanical coupling rate  must consider both effects, $g_0=g_\text{pe}+g_\text{mb}$, where $g_\text{pe}$ stands for the volumetric photo-elastic contribution (\textit{pe}-effect), and $g_\text{mb}$ for the cavity moving-boundaries contribution (\textit{mb}-effect). We focus on the mechanical modes that are phase-matched with fundamental TM (tranverse magnetic) optical mode (\cref{figure4}c) since the TM-modes exhibits the largest coupling rate and potentially higher optical quality factors~\cite{Borselli:2004ds}.  In ~\cref{figure4}a,b we show the photo-elastic ($g_\text{pe}$,red), the moving-boundary ($g_\text{mb}$, green) and total coupling rate $g_0$ (bars) for the phase-matched mechanical modes in the single (\cref{figure4}a) and double-disk (\cref{figure4}b) structures. 
\tocless{\subsection*{Optomechanical coupling}}
For both $sd$ and $dd$-structures, the whispering,  Rayleigh, and dilatational and mode groups can be identified in \cref{figure4}a,b. The relative contributions from the \textit{pe} and \textit{mb}-effects however varies significantly for each structure and mode group. To understand this in detail, we analyze the weighting function role played by the optical field in the $mb$ and $pe$-effects (see Methods). For the $mb$-effect, the optical  weighting of the normal mechanical displacement ($u_\bot$) along the radially parallel cavity boundaries is given by (see Supplementary Information),
	\begin{equation}
		\rho_\text{mb}=\delta\epsilon_\text{mb}E_{r}^2 - \delta\epsilon_\text{mb}E_{\phi}^2-\delta\epsilon_\text{mb}^{-1}D_z^2,
		\label{eq:gmb_terms}
	\end{equation}	
where $E_{r}$, $E_{\phi}$ and $E_{z}$ are the energy-normalized electric field components, $\delta\epsilon_{\text{mb}}=\epsilon_1-\epsilon_2$ and $\delta\epsilon_{\text{mb}}^{-1}=(1/\epsilon_1-1/\epsilon_2)$ with $\epsilon_1=\epsilon_{0}n^{2}_{1}$  and $\epsilon_2=\epsilon_{0}n^{2}_{2}$ being the permittivities of the silicon and air, respectively. The spatial dependence of the three terms in eq. \ref{eq:gmb_terms} are shown in \cref{figure4}d,e for both $sd$ and $dd$-cavities. It is evident that the $mb$-effect is dominated by the azimuthal component $-\delta\epsilon_\text{mb}|E_{\phi}|^2$ in both structures. The opposite sign of the azimuthal term relative to the radial contribution \textemdash~due to the $\pi$ phase difference between forward and backward azimuthal field components \textemdash~drastically distinguishes the backward from the forward Brillouin optomechanical interaction. The peaking around $r=4.6$~\textmu m of the weighting terms also hints which mechanical modes should benefit from the $mb$-effect. 
	As for the \textit{pe}-effect contribution ($g_\text{pe}$) is mostly due to the anisotropic permitivitty components $\delta\epsilon_\text{pe}^{zz}$ and $\delta\epsilon_\text{pe}^{\phi\phi}$ for TM optical mode; the anisotropic components are calculated from the permitivitty perturbation tensor, defined as $\delta\vec{\epsilon}_{\text{pe}}=-\epsilon_{0}\,n^4_{1}\,\vec{p}\mkern1mu{:}\vec{S}$, in which $\vec{p}$ is the photoelastic tensor of the isotropic silicon and $\vec{S}$ is the strain tensor induced by the mechanical waves (see Methods). 
 In silicon, the dominant photo-elastic coefficient ($p_{11}=-0.09,p_{12}=0.017$) is $p_{11}$ and therefore an insight about which modes will lead to a strong $pe$-effect can be obtained using ($\delta\epsilon_\text{pe}^{\phi\phi}\approx -\epsilon_{0}\,n_{1}^{4}\,p_{11}\,S_{\phi\phi}$) and vertical ($\delta\epsilon_{\text{pe}}^{zz}\approx -\epsilon_{0}\,n_{1}^{4}\,p_{11}\,S_{zz}$). 

The $w$-mode group has the largest optomechanical coupling rate and give rise to several peaks in the high frequency range ($20\sim 27$ GHz), which are unique to BBS due to the large mechanical azimuthal numbers imposed by the phase-matching condition. 
Due to their dominant displacement components ($u_r,u_\phi$), their largest strain component is along the azimuthal direction $S_{\phi\phi}\approx M u_\phi/r$. Such a large azimuthal strain lead to a $pe$-effect dominated optomechanical coupling, reaching the highest coupling rate for the single-disk, $g_0/2\pi\approx 61$ kHz at $24.34$~GHz for the 16th radial order $w$-mode ($w_{16}$), whose profile is show in (\cref{figure4}f). In this mode group $\delta\epsilon_\text{pe}^{\phi\phi}$ accounts for about 70\% of the total coupling coefficient (\cref{table:1}, see Supplementary Information). The tiny in-plane displacement ensures small contribution from the $mb$-effect.
The peaked  $g_0$ distribution, which could be anticipated by the fine frequency spacing  for this mode family can be understood by inspecting the overlap between the dominant azimuthal strain component $S_{\phi\phi}$ and the optical mode profile.  Despite the strain oscillations along the radial direction, there is a net tensile strain ($S_{\phi\phi}>0$) region indicated by the dashed arrow in (\cref{figure4}f). As the frequency increases, the net strain region shifts inwards along the disk and sweeps the spatial matching between the strain and optical fields. Underlying the existence of this net positive strain region is the hybrid longitudinal-transverse nature of the $w$-group, which can be precisely traced using the analytic solution of an infinite cylinder: the fast radial oscillation periods seen in (\cref{figure4}f) arise from the transverse-wave contribution to this mode, whereas the slower net positive strain is caused by  longitudinal-wave contribution (see Supplementary Information).

The Rayleigh mode ($r_1$), which has the lowest resonant frequency  (at 11.12~GHz, \cref{figure4}a,g), has a dominant radial displacement ($u_r$) in the single-disk structure, whereas the vertical ($u_z$) component is dominant for its odd flexural-like counterpart in the double-disk structure. 
The strain $S_{\phi z}$ \textemdash~shown in \cref{figure4}g \textemdash~ is the dominant strain for $f_{1}^{dd}$ mode.
 In the \textit{sd}-cavity, the minor role of the $mb$-effect is expected as the boundary deformation is concentrated at the disk edge, while the optical field components are localized at the disk's top and bottom surface (see \cref{figure4}c). Indeed, the fundamental $f_1^\text{dd}$ mode at 11.15~GHz has the second largest coupling rate among  all the $dd$-cavity modes, reaching $g_0/2\pi=31$~kHz. The $dd$-cavity further allows tailoring of the \textit{mb}-effect strength  by adjusting the slot height between the two disks, in \cref{figure4}i we show that the total coupling rate ($g_0$) of the $f_1^\text{dd}$ mode can be improved by $300\%$ by reducing the slot height from $150$~nm to $50$~nm.

Finally, a very high  optomechanical coupling rate could be expected for the $d$-mode group ($16<\Omega/2\pi<18$~GHz). These modes display not only a large vertical strain but also a large deformation of the cavity boundaries, as shown in \cref{figure4}h. Indeed these two effects are very strong individually but their opposite sign lead to a cancellation effect, a clear competition between the \textit{mb} ($g_\text{mb}$) and \textit{pe}-effects ($g_\text{mb}$)~\cite{Florez:2016aa}. For the double-disk structure, whose optical weighting function is shown in  \cref{figure4}e, the slot effect enhancement does not readily improve the optomechanical coupling with the dilatational modes, this is due to a balanced contribution from the azimuthal (dashed black line) and vertical field (dashed blue line) components, which oppositely contribute to the  \textit{mb}-effect and almost cancel it.
\tocless{\subsection*{Discussion}}
Using the calculated mechanical frequencies and BBS coupling rates for the $sd$ and $dd$-cavity designs we can estimate the power threshold for the stimulated Brillouin lasing. Assuming an 1550 nm optical mode and  conservative optical and mechanical mode parameters, intrinsic optical quality factor of $2\times 10^5$ ($\kappa_\text{p}/2\pi=\kappa_\text{s}/2\pi\approx 1.2$ GHz), mechanical quality factor of $10^3$, and simultaneous resonance condition for both pump and stokes wave ($\Delta_\text{s}=\Delta_\text{p}=0$), \cref{eq:threshold} predicts a threshold of only  $P_\text{th}\approx(8;31)$ mW for the  ($w_{16}$,w$^{dd}_{16}$) modes. For cantilever-like flexural mode of the double-disk ($f_{1}^{dd}$), the threshold power is $P_\text{th}\approx 17$ mW (assuming the same mechanical quality factor). The threshold power for the$f_{1}^{dd}$-mode can be reduced even further for smaller gaps. For instance, if  $t_{\text{SiO}}=50$~nm is possible to achieve $g_{0}/2\pi\approx 75$~kHz (\cref{figure4}i), leading to a threshold power of only $P_\text{th}\approx 3$~mW (considering the same optical and mechanical losses). Experimentally, in order to ensure the simultaneous resonant condition, a set of coupled optical cavities with varying coupling gaps could be fabricated. The importance of the compound cavity scheme becomes clear if we compare the single-resonance threshold, which is predicted by \cref{eq:threshold} assuming a resonant stokes signal $\Delta_\text{s}=0$ and the optimal pump-detuning ($\Delta_\text{p}=\Omega_\text{m}$). Using the same optical and mechanical linewidth above  and a single-photon optomechanical coupling rate ($g_0 = 2 g_0^c$), the threshold for Brillouin lasing in a single-resonance scheme would be as high as  $P_\text{th}\approx 3.2$~W for the $w_{16}$ mode, which is impractical due to strong detrimental effects, such as nonlinear light absorption in silicon. 
\tocless{\section*{Conclusions}}	
Our results provides a clear guideline towards the observation of stimulated backward Brillouin scattering in an integrated CMOS-compatible silicon device. The results are promising even for compound cavities based on standard single-disk silicon devices. The double-disk device, although exhibiting a lower optomechanical coupling, may benefit from the potentially higher mechanical quality factor of the lower-frequency cantilever-like mechanical modes. Our findings  indicate that coupled silicon single and double-disk resonators offer large optomechanical coupling rates an the necessary degrees of freedom to engineer and manipulate the Brillouin scattering in compact structures. Although we concentrated our discussion on silicon-based devices, our results could be adapted to similar structures fabricated from other  high index materials, such as III-V, Si$_3$N$_4$ and SiO$_2$.

\tocless{\section*{Methods}}

\textbf{Frequency splitting} The frequency splitting simulation was performed using a two-dimensional approximation to the actual \textit{sd}-cavity. In this approximation, the modes of the coupled infinite cylinders are calculated while constraining the out-of-plane wave number, 
\begin{equation}
k_{z}=\sqrt{(k_{0}n)^2-\left(\frac{m}{r}\right)^2-\left(\frac{z_{m,1}}{r}\right)^2},
\label{eq:kzoutp}
\end{equation}
where $m/r$ and $z_{m,1}/r$ are the azimuthal and radial components of the wave vector with norm $k_{0}\,n$, $k_{0}$ is the free-space wave number for $\lambda=1.55$~\textmu m, $n=3.5$ is the silicon refractive index, $r=5$~\textmu m, $m=35$ is the optical azimuthal number for the TM-mode of the \textit{sd}-cavity and $z_{m,1}$ is the first zero of the Bessel function $J_{m}(z)$. This is equivalent to the Marcatilli effective index method and we verified that it leads to a electric field envelope that agrees well with the numerical mode obtained from the axisymmetric calculation.

\textbf{Mechanical mode dispersion} We obtain the dispersion relation $\Omega(M)$ by solving the eigenfrequency problem derived from the full-vectorial elastic wave equation through the finite-element method (see Supplementary Information), due to the highly multimode character of the mechanical dispersion, we show in \cref{figure3}a-c a grey color shading proportional to the mechanical density of modes instead of the calculated pairs $(\Omega_M,M)$. The mechanical density of modes is calculated as $\rho(M,\Omega)=\sum_{i,j}f(M,M_{0}^{(i)},\sigma_{\text{M}})\,g(\Omega,\Omega_{0}^{(ij)},\sigma_{\Omega})$, where $f$ and $g$ are Gaussian weighting functions with a normalized product, full-width-half-maximum (FWHM) wavenumber $\sigma_\text{M}=0.1$, FWHM-frequency $\sigma_{\Omega}/2\pi=117.5$~MHz and given one azimuthal acoustic number $M_{0}^{(i)}$ are calculated each of the frequencies $\Omega_{0}^{(ij)}$ from the elastic wave equation.

\textbf{Coupled mode equations} The resulting energy exchange between the optical pump and Stokes wave can be modeled using coupled mode theory \cite{Matsko:2002vs,Agarwal:2013hl} (see Supplementary Information), which leads to a set of coupled equations for the amplitudes of the pump wave ($a_\text{p}$), stokes wave ($a_\text{s}$) and mechanical wave ($b$),
\begin{eqnarray} \label{eq:coupled_eqs}
\nonumber
\dot{a}_\text{p}&=&(i\,\Delta_\text{p}\,-\kappa_\text{p}/2)a_{\text{p}} -\,i\,g_{0}^\text{c}\,b\,a_{\text{s}}+\sqrt{\kappa_{\text{e}}} s_{\text{p}},\\ 
\dot{a}_\text{s}&=&(i\,\Delta_\text{s}\,-\kappa_\text{s}/2)a_{\text{s}}-\,i\,g_{0}^\text{c}\,b^{*}\,a_\text{p},\\ \nonumber
\dot{b}&=&(-i \Omega_\text{m}-\Gamma_\text{m}/2)b-i\,g_{0}^\text{c}\,a_\text{p}\,a_\text{s}^{*}+F_\text{th},
\end{eqnarray}
where $a_\text{i}$'s are normalized such that $|a_\text{i}|^2$ is the intra-cavity photon number and $b$ is normalized such that $|b|^{2}$ is the phonon number. $\Delta_\text{i}=\omega_\text{i}-\omega_{\text{i}_0}$ is the frequency detuning between the pump ($\text{i=p}$) or stokes wave ($\text{i=s}$) relative to the optical cavity mode frequencies $\omega_{\text{p}_0,\text{s}_0}$. 
$\kappa_\text{p,s}$ represents the optical loss rate for each mode, ($\Omega_{\text{m}},\Gamma_\text{m}$) are the mechanical mode frequency and damping rate, respectively; the optomechanical coupling rate  is $g_0^\text{c}$ and represents the photon coupling rate between the stokes and the pump wave induced by the zero-point fluctuation of the mechanical mode, which will be calculated shortly using electromagnetic perturbation theory~\cite{Haus:1991aa,Johnson:2002tp}. Note that $g_0^\text{c}$ is related to the more usual single-cavity coupling rate as $g_0^\text{c}=g_0/2$. The factor $1/2$ arises because the coupled optical mode is distributed within two optical cavities whereas the mechanical mode is localized within a single cavity due to the air gap between the cavities. $\kappa_\text{e}$ is extrinsic coupling rate to the feeding waveguide carrying a photon-flux $|s_\text{p}|^2$. $F_\text{th}$ is a white-noise random thermal force responsible for driving the mechanical motion.

\textbf{$g_0$ calculation} The \textit{mb}-effect contribution is given by~\cite{Johnson:2002tp}(see Supplementary Information), 
\begin{equation}
g_\text{mb} =-\frac{\omega_\text{p}}{2} \oint_{S} u_\bot(\delta\epsilon_\text{mb} \vec{E}_{\text{p},\parallel}^{*} \mkern1mu{\cdot} \vec{E}_{\text{s},\parallel} -\delta\epsilon^{-1}_\text{mb}\vec{D}_{\text{p},\bot}^{*}\mkern1mu{\cdot}\vec{D}_{\text{s},\bot})dA,\label{eq:g_mb}	
\end{equation}
where the permittivity differences are given by $\delta\epsilon_{\text{mb}}=\epsilon_1-\epsilon_2$ and $\delta\epsilon_{\text{mb}}^{-1}=(1/\epsilon_1-1/\epsilon_2)$, in which $\epsilon_1=\epsilon_{0}n^{2}_{1}$  and $\epsilon_2=\epsilon_{0}n^{2}_{2}$ are the permittivities of the silicon and air, respectively. $u_\bot=x_\text{zpf}\, \vec{u}\cdot\hat{n}$ is the surface-normal component of the displacement vector $\vec{u}$ (normalized to unit); $x_\text{zpf}=\sqrt(\hbar/2 m_\text{eff}\Omega)$ is zero-point fluctuation of the mechanical mode with effective mass $m_\text{eff}$; the fields $\vec{\vec{E}}_{j,\parallel}$ and $\vec{\vec{D}}_{j,\bot
}$ are boundary-tangential electric field and boundary-normal electric displacement field to the cavity surface $S$ of the pump ($j=\text{p}$) or scattered ($j=\text{s}$) optical mode (energy-normalized). The \textit{pe}-effect contribution is given by~\cite{Haus:1991aa}(see Supplementary Information),
\begin{equation}
g_\text{pe}=-\frac{\omega_\text{p}}{2}\int_{V} \vec{\vec{E}}_\text{p}^{*} \mkern1mu{\cdot} \delta\vec{\epsilon}_\text{pe} \mkern1mu{\cdot}\vec{\vec{E}}_\text{s}\,dV,
\label{eq:g_pe}
\end{equation}
where $\delta\vec{\epsilon}_{\text{pe}}=-\epsilon_{0}\,n^4_{1}\,\vec{p}\mkern1mu{:}\vec{S}$ is the photo-elastic perturbation in the permittivity inside the cavity volume $V$, $\vec{p}$ is the photoelastic tensor of silicon, and $\vec{S}=x_\text{zpf}\nabla_s \vec{u}$ is the strain tensor induced by the mechanical waves.  The optical and elastic mode profiles are numerically calculated using the finite-element method (see Supplementary Information).

\textbf{Simulation parameters} Si refractive index $n_\text{Si}=3.5$, silica refractive index $n_{\text{SiO}_{2}}=1.45$, air refractive index $n_{\text{air}}=1.0$, wavelength of interest $\lambda = 1550$~nm, Si Young's modulus $E_{\text{Si}}=170$~GPa, silica Young's modulus $E_{\text{SiO}_{2}}=72$~GPa, Si Poisson's ratio $\nu_{\text{Si}}=0.28$, silica Poisson's ratio $\nu_{\text{SiO}_{2}}=0.17$, Si density mass $\rho_{\text{Si}}=2329$~kg/m$^3$, silica density mass $\rho_{\text{SiO}_{2}}=2203$~kg/m$^3$ and Si photo-elastic coefficients $p_{11}=-0.09$, $p_{12}=0.017$ and $p_{44}=-0.0535$. Here we neglect silicon anisotropy and assume the values of $E_{\text{Si}}$ and $\nu_{\text{Si}}$ along to principal crystal axes~\cite{hopcroft2010young}. The silicon photoelastic coefficients used in the simulations are taken from ref.~\cite{Biegelsen:1974}.

\textbf{Acknowledgements} The authors would like to acknowledge Omar Florez and Paulo Dainese for fruitful discussions. This research was funded by the Sao Paulo State Research Foundation (FAPESP) (Grants 2012/17765-7 and 2012/17610-3), the National Counsel of Technological and Scientific Development (CNPQ), and the Coordination for the Improvement of Higher Education Personnel (CAPES).

\textbf{Author contribution} Y.E. performed  the numerical simulations and conceived the idea with help from G.W. and T.A. G.O.L and F.S helped Y.E. with the numerical simulation and analytical analysis. All authors discussed the results and their implications and contributed to writing this manuscript.

\textbf{Competing financial interest} The authors declare that they have no competing financial interests.

\renewcommand{\theequation}{S\arabic{equation}}
\renewcommand{\thesection}{S\arabic{section}}
\renewcommand{\thesubsection}{\Alph{subsection}}
\renewcommand{\thesubsubsection}{\roman{subsubsection}}
\renewcommand{\thefigure}{S\arabic{figure}}
\renewcommand{\thetable}{S\arabic{table}}
\setcounter{figure}{0}
\setcounter{table}{0}
\setcounter{equation}{0}
\setcounter{section}{0}
\onecolumngrid
\newpage

\tocless{\section*{Supplementary Information}}
\tableofcontents
\section{\label{sup:cpm}Coupled mode equations}
We derive the coupled mode equations for the optical and mechanical modes following an approach similar to \cite{Agarwal:2013hl}. The electric field is obtained from Maxwell's wave equation in the presence of a time-dependent polarization term,  
\begin{equation}
\nabla\times\nabla\times\vec{\mathcal{E}}=-\mu_{0}\epsilon\partial^{2}_{t} \vec{\mathcal{E}}-\mu_{0}\partial^{2}_{t}(\delta\vec{P}),
\label{eq:max1}
\end{equation}
where $\vec{\mathcal{E}}$ is total electric field vector, $\mu_0$ is the vacuum permeability, $\epsilon$ is the isotropic unperturbed spatial permittivity. The additional polarization, $\delta\vec{P}$, arises from the mechanical mode perturbation to the optical field. The mechanical modes are described by the equation of motion,
\begin{equation}
\nabla\mkern1mu{\cdot}\left(\vec{c}\mkern1mu{:}\vec{\mathcal{S}}\right)-\rho\,\partial_t^2\,\vec{\mathcal{U}}=-\vec{\mathcal{F}},
\label{eq:sme1}
\end{equation}
where $\vec{\mathcal{U}}$ is the mechanical displacement, $\vec{c}$ is the stiffness tensor, $\vec{\mathcal{S}}=\nabla_{\text{s}}\,\vec{ \mathcal{U}}$ is the strain tensor, $\rho$ is the material density, and $\vec{\mathcal{F}}$ is the force density vector with contributions from the electric part of the Maxwell stress tensor and electrostriction tensor~\cite{Boyd:2050257}. 

To obtain the coupled mode equations for the optical fields we expand $\vec{\mathcal{E}}$ in terms of slowly-varying amplitudes for the pump (p) and stokes (s) fields, we consider the modal expansion,
\begin{equation}
\vec{\mathcal{E}}(\vec{r},t)=\sum_{j=\text{p},\text{s}}a_j(t)e^{-i\,\omega_{j}\,t}\vec{E}_{j}(\vec{r})+c.c.
\label{eq:cpm1}
\end{equation}
The optical mode spatial distribution $\vec{E}_{j}(\vec{r})$ is normalized such that $\sum_{j} |a_{j}|^2$ represents the total optical energy. Each modal fields satisfy the Helmholtz equation,
\begin{equation}
\nabla\times\nabla\times\vec{E}_{j}=\omega_{0,j}^{2}\,\mu_{0}\,\epsilon\vec{E}_{j},
\label{eq:max2}
\end{equation}
where $\omega_{0,j}$ is the resonant frequency of each optical mode. These modes are orthonormalized,
\begin{equation}
\int\vec{E}_m^{*}\mkern1mu{\cdot}\epsilon\vec{E}_{n}\,dV=\delta_{m,n}.
\label{eq:enorm}
\end{equation}
Substituting \cref{eq:cpm1} in \cref{eq:max1}, exploring the slowly-varying envelope approximation (SVEA) ($d/dt\ll \omega_{j}$) and the small detuning  approximation, $\omega_{j}^{2}-\omega_{0,j}^{2}\approx 2\omega_{j}  \Delta_{j} $ (with $\Delta_j=\omega_j-\omega_{0,j}$) we arrive at the following coupled equations for the field amplitudes $a_j$,
\begin{equation}
\sum_j \left[2\omega_{j}\left(i\dot{a}_j+\Delta_j a_j\right)\right]e^{-i\omega_{j}t}\epsilon\vec{E}_j+c.c.=\partial_{t}^{2}(\delta\vec{P}).
\label{eq:cpm2}
\end{equation}
We can decouple \cref{eq:cpm2} by multiplying by $\vec{E}_l^{*}$ in both sides of \cref{eq:cpm2}, integrating over the whole space, and using \cref{eq:enorm},

\begin{equation}
\left(i\dot{a}_l+\Delta_l a_l\right)e^{-i\omega_{l}t}+c.c.=\frac{\int[\vec{E}_{l}^{*}\mkern1mu{\cdot}\partial_{t}^{2}(\delta\vec{P})]dV}{2\omega_{l}}.
\label{eq:cpm3}
\end{equation}
The spatial and time-dependence of the polarizability  is given by,
\begin{equation}
\delta\vec{P}(\vec{r},t) =\delta\boldsymbol\varepsilon(\vec{r},t)\mkern1mu{\cdot}\vec{E}(\vec{r},t),
\label{eq:deltaP}
\end{equation}
where the time-dependence of the permittivity perturbation $\delta\boldsymbol\varepsilon(\vec{r},t)$ will be given by time-dependence of the mechanical mode, 

\begin{equation}
\vec{\mathcal{U}}(\vec{r},t)=b(t)e^{-i\Omega t}\vec{u}(\vec{r})+c.c.,
\label{eq:mec_disp}
\end{equation}
where we choose to normalize the mechanical spatial distribution such that $\max(|\vec{u}(\vec{r})|)=1$, therefore $b(t)$ has units of length. 

The mechanical mode will perturb the optical mode both through the photo-elastic (\textit{pe}) effect and through of the moving boundary (\textit{mb}) effect. Either contributions will be proportional to the displacement amplitude $b(t)$. Therefore the permittivity perturbation time-dependence can be factored out as $\delta\boldsymbol\varepsilon(\vec{r},t)=(b(t)/b_{0})\exp(-i\Omega t)\delta\boldsymbol\epsilon(\vec{r})+c.c.$, where $\delta\boldsymbol\epsilon(\vec{r})$ is the spatial permittivity perturbation and $b_{0}$ is a free-parameter of amplitude normalization with units of length. Substituting this expression together with \cref{eq:cpm1} into \cref{eq:deltaP} we obtain,

\begin{equation}
\delta\vec{P}(\vec{r},t) = \left(\frac{1}{b_{0}}\right)\left(b(t)e^{-i\Omega t}\delta\boldsymbol\epsilon(\vec{r})+c.c\right)\mkern1mu{\cdot}\left(\sum_{m}a_{m}(t)e^{-i\omega_m t}\vec{E}_{m}(\vec{r})+c.c.\right).
\label{eq:deltaP_prod}
\end{equation}

There will be four distinct terms for each term $m$ in the summation \cref{eq:deltaP_prod},

\begin{align*}
&b(t)a_{m}(t)\delta\boldsymbol\epsilon(\vec{r})\mkern1mu{\cdot}\vec{E}_{m}(\vec{r})e^{i(-\omega_{m}-\Omega)t}+
b^{*}(t)a_{m}^{*}(t)\delta\boldsymbol\epsilon^{*}(\vec{r})\mkern1mu{\cdot}\vec{E}_{m}^{*}(\vec{r})e^{i(\omega_{m}+\Omega)t}\\
+&b(t)a_{m}^{*}(t)\delta\boldsymbol\epsilon(\vec{r})\mkern1mu{\cdot}\vec{E}_{m}^{*}(\vec{r})e^{i(\omega_{m}-\Omega)t}
+b^{*}(t)a_{m}(t)\delta\boldsymbol\epsilon^{*}(\vec{r})\mkern1mu{\cdot}\vec{E}_{m}(\vec{r})e^{i(-\omega_{m}+\Omega)t}.
\end{align*}

In a rotating-wave approximation (RWA) these distinct terms will be relevant drives to the \cref{eq:cpm3} provided they satisfy the energy conservation. This will depend whether we are treating the pump ($a_\text{p}$) or Stokes ($a_\text{s}$) amplitudes. For example, for the Stokes wave $\omega_\text{s}=\omega_\text{p}-\Omega$ only the last term is relevant. The time-derivative of the polarization in \cref{eq:cpm3} will have terms involving the first and second derivatives of the slowly varying amplitudes $a_{m}(t),b(t)$ and terms of the order of $\omega_m^2$. Employing the SVEA and choosing the relevant driving terms from \cref{eq:deltaP_prod}, we can finally write the amplitude equations for \textit{positive frequency} amplitudes of the optical fields; dropping out the fast oscillating terms in both sides of \cref{eq:cpm3} lead to,
\begin{align}
\dot{a}_\text{p} &= i\,\Delta_{\text{p}}\,a_{\text{p}}-\frac{i\,g_{0}\,b\,a_{\text{s}}}{b_{0}},\label{eq:CMEopt1}\\
\dot{a}_\text{s} &= i\,\Delta_{\text{s}}\,a_{\text{s}}-\frac{i\,g_{0}^{*}\,b^{*}\,a_{\text{p}}}{b_{0}},\label{eq:CMEopt2}
\end{align}
where, 
\begin{equation}
g_{0} = -\frac{\omega_\text{p}}{2}\int_{V}\vec{E}_{\text{p}}^{*}\mkern1mu{\cdot}\delta\boldsymbol\epsilon\mkern1mu{\cdot}\vec{E}_{\text{s}}\,dV, 
\label{coucoef}
\end{equation}
represents the optomechanical coupling rate and describes the frequency shift of the pump wave generated by the scattering from an acoustic wave, with an amplitude equivalent to the zero-point fluctuation ($x_{\text{zpf}}$), that perturb the dielectric constant by $\delta\boldsymbol\epsilon$.

To find the equation of motion for the mechanical mode we proceed in a similar fashion. Each mechanical mode satisfies the modal equation,
\begin{equation}
\nabla\mkern1mu{\cdot}\left(\vec{c}\mkern1mu{:}\vec{S}\right)=-\rho\,\Omega_{0}^{2}\,\vec{u},
\label{eq:mech_eigenmode}
\end{equation}
where $\vec{S}=\nabla_{\text{s}}\,\vec{u}$ is the spatial distribution of the strain tensor per unit length and $\Omega_{0}$ is the mechanical mode resonant frequency. Substituting the mechanical mode expansion \cref{eq:mec_disp} in \cref{eq:sme1} and, in the resulting equation, substituting \cref{eq:mech_eigenmode} and exploring the small-detuning approximation $\Omega^2-\Omega_{0}^2\approx 2\,\Omega\,\Delta_\text{m} $ (with $\Delta_\text{m}=\Omega-\Omega_{0}$), we arrive at,

\begin{equation}
2\,\Omega(i\dot{b}+\Delta_\text{m} b)e^{-i\,\Omega\,t}+c.c.=\frac{\left<\vec{u}|\vec{\mathcal{F}}\right>}{m_\text{eff}},
\label{eq:disp_cpm}
\end{equation}
where $\left<\vec{u}|\vec{\mathcal{F}}\right> = \int\vec{u}^{*}\mkern1mu{\cdot}\vec{\mathcal{F}}\,dV$, $m_\text{eff}=\int\rho|\vec{u}|^{2}dV$ is the effective motional mass, $\vec{\mathcal{F}}=\vec{\mathcal{F}}_{\text{MT}}+\vec{\mathcal{F}}_{\text{ES}}$, in which $\vec{\mathcal{F}}_{\text{MT}}=\nabla\mkern1mu{\cdot}\vec{\mathcal{T}}$ is the force density from the Maxwell stress tensor and $\vec{\mathcal{F}}_{\text{ES}}=-\nabla\mkern1mu{\cdot}\boldsymbol\varsigma$ is the force density from the electrostriction tensor,
\begin{align}
\mathcal{T}_{ij} &= \epsilon\left(\mathcal{E}_i\,\mathcal{E}_j -\frac{1}{2}\delta_{ij}|\vec{\mathcal{E}}|^2\right),\label{eq:stress_MT}\\
\varsigma_{ij} &= \gamma_{ijkl}\,\mathcal{E}_k\,\mathcal{E}_l,\label{eq:stress_ES}
\end{align}
are the time-dependent electric Maxwell and electrostriction stress tensors, respectively, $\gamma_{ijkl}=-(1/2)\epsilon_{0}n^{4}p_{ijkl}$, where $n$ being the optical refractive index and $p_{ijkl}$ being the photoelastic tensor. The minus sign used in the definition of the electrostrictive force  follows the conservative force convention~\cite{Boyd:2050257,panofsky2012classical}. Using our field expansion \cref{eq:cpm1}, the general form of the field products in \cref{eq:stress_MT} and \cref{eq:stress_ES},


\begin{equation}
\mathcal{E}_i\mathcal{E}_j=\left(\sum_{l} a_l(t)e^{-i\omega_l t}E^{(i)}_{l}+c.c.\right)\left(\sum_{m} a_m(t) e^{-i\omega_m t}E^{(j)}_{m}+c.c.\right),
\label{eq:field_product}
\end{equation}
where the parenthesis superscript $(i,j)$ indicate the spatial component of the modal field. According to RWA, among all terms in \cref{eq:field_product} the only relevant ones are those oscillating at the mechanical frequency, i.e., terms with frequencies $\omega_{\text{p}}-\omega_{\text{s}}$. Therefore, substituting \cref{eq:field_product} in \cref{eq:stress_MT} and \cref{eq:stress_ES}, considering the relevant terms,


\begin{align}
\mathcal{T}_{ij} &= a_{\text{p}}\,a_{\text{s}}^{*}\,T_{ij}\,e^{-i(\omega_{\text{p}}-\omega_{\text{s}})t}+c.c.,\label{stress2b}\\
\varsigma_{ij} &= a_{\text{p}}\,a_{\text{s}}^{*}\,\sigma_{ij}\,e^{-i(\omega_{\text{p}}-\omega_{\text{s}})t}+c.c.,\label{stress2a}
\end{align}
where
\begin{gather}
T_{ij} = \epsilon[E_{\text{p}}^{(i)}E_{\text{s}}^{(j)*}+E_{\text{p}}^{(j)}E_{\text{s}}^{(i)*}-\delta_{ij}\vec{E}_{\text{p}}\mkern1mu{\cdot}\vec{E}_{\text{s}}^{*}],
\label{stress3}\\
\sigma_{ij} = \gamma_{ijkl}[E_{\text{p}}^{(k)}E_{\text{s}}^{(l)*}+E_{\text{p}}^{(l)}E_{\text{s}}^{(k)*}],\label{stress33}
\end{gather}
are the spatial distributions of the electric Maxwell and electrostriction  stress tensors, respectively. Therefore, substituting \cref{stress2b} and \cref{stress2a} in the driving term,
\begin{equation}
\left<\vec{u}|\vec{\mathcal{F}}\right> = \int \vec{u}^{*}\mkern1mu{\cdot}\nabla\mkern1mu{\cdot}\vec{\mathcal{T}}\,dV-\int \vec{u}^{*}\mkern1mu{\cdot}\nabla\mkern1mu{\cdot}\boldsymbol\varsigma\,dV,
\label{eq:mst_es}
\end{equation}
and substituting the resulting equation in \cref{eq:disp_cpm} and then time-averaging, 
\begin{equation}
\dot{b}=i\,\Delta_{\text{m}}\,b+\frac{i\,a_{\text{p}}\,a_{\text{s}}^{*}}{2\,\Omega}\frac{\left<\vec{u}|\vec{f}\right>}{m_{\text{eff}}},
\label{mampeq}
\end{equation}
where
\begin{equation}
\left<\vec{u}|\vec{f}\right> = \int\vec{u}^{*}\mkern1mu{\cdot}\vec{f}\,dV = \int\vec{u}^{*}\mkern1mu{\cdot}(\nabla\mkern1mu{\cdot}\mathbf{T}-\nabla\mkern1mu{\cdot}\boldsymbol\sigma)\,dV. \label{overlapa}
\end{equation}

It is known that the electric Maxwell stress tensor lead only to a boundary force in a transparent material, whereas the electrostriction tensor leads to a volume force~\cite{Rakich:2012et,Wolff:2015ji}. These two contributions can be obtained by integrating by parts \cref{overlapa} and disregarding the electrostriction surface pressure term~\cite{Wolff:2015ji,panofsky2012classical},

\begin{equation}
\left<\vec{u}|\vec{f}\right> = \oint_{S}\vec{u}^{*}\mkern1mu{\cdot}\vec{f}_{\text{rp}}\,dA+\int_{V}\boldsymbol\sigma\mkern1mu{:}\vec{S}^{*}\,dV,
\label{eq:driveneq}
\end{equation}
where the double inner product is defined as $\boldsymbol\sigma\mkern1mu{:}\vec{S}^{*}=\sigma_{ij}S_{ij}^{*}$, and,

\begin{equation}
\vec{f}_{\text{rp}}=(\mathbf{T}_{2}-\mathbf{T}_{1})\mkern1mu{\cdot}\hat{n},
\label{eq:RPeq}
\end{equation}
represent the spatial distribution of the radiation pressure making on the surface $S$ of the cavity with volume $V$. $\mathbf{T}_{1}$ and $\mathbf{T}_{2}$ are the Maxwell stress tensors calculated inside and outside of the cavity, respectively, and $\hat{n}$ is the unitary normal vector to $S$ that points from inside to outside of the cavity.\\ 


A more convenient form of the radiation pressure is obtained when \cref{stress3} is substituting in \cref{eq:RPeq} considering two different materials and continuous fields on $S$, 
 
\begin{equation}
\vec{f}_{\text{rp}}=[\delta\epsilon_{\text{mb}}(\vec{E}^{*}_{\text{s},\parallel}\mkern1mu{\cdot}\vec{E}_{\text{p},\parallel})
-\delta\epsilon_{\text{mb}}^{-1}(\vec{D}^{*}_{\text{s},\perp}\mkern1mu{\cdot}\vec{D}_{\text{p},\perp})]\hat{n},
\label{eq:RP_PS}
\end{equation}
in which $\vec{E}_{p,\parallel}$ and $\vec{E}_{s,\parallel}$ are the parallel electric fields from pump and the Stokes waves, respectively, $\vec{D}_{p,\perp}$ and $\vec{D}_{s,\perp}$ are the perpendicular electric displacements from pump and Stokes waves, respectively, $\delta\epsilon_{\text{mb}}=\epsilon_{0}(n_{1}^{2}-n_{2}^{2})$ and $\delta\epsilon_{\text{mb}}^{-1}=\epsilon_{0}^{-1}(n_{1}^{-2}-n_{2}^{-2})$. As parallel electric fields and perpendicular electric displacements are normalized then the radiation pressure is also normalized such that it has units of $L^{-3}$, as can be evaluated from \cref{eq:RP_PS}.\\

Like radiation pressure, second term in the right-hand of \cref{eq:driveneq} can be also written in a more convenient form,


\begin{equation}
\int_{V}\boldsymbol\sigma\mkern1mu{:}\vec{S}^{*}\,dV = \int_{V}(\vec{E}_{\text{p}}^{*}\mkern1mu{\cdot}\delta\boldsymbol\epsilon_{\text{pe}}\mkern1mu{\cdot}\vec{E}_{\text{s}})^{*}\,dV,\label{eq:ESdriven1}
\end{equation}
where $\delta\boldsymbol\epsilon_{\text{pe}}=-\epsilon_{0}\,n^{4}\,\vec{p}\mkern1mu{:}\vec{S}$ is the anisotropic perturbation in the permittivity per unit length from the photoelastic effect. Now substituting \cref{eq:RP_PS,eq:ESdriven1} in \cref{eq:driveneq} results,

\begin{equation}
\left<\vec{u}|\vec{f}\right> = \oint_{S}(\vec{u}^{*}\mkern1mu{\cdot}\hat{n})[\delta\epsilon_{\text{mb}}(\vec{E}^{*}_{\text{s},\parallel}\mkern1mu{\cdot}\vec{E}_{\text{p},\parallel})
-\delta\epsilon_{\text{mb}}^{-1}(\vec{D}^{*}_{\text{s},\perp}\mkern1mu{\cdot}\vec{D}_{\text{p},\perp})]dA+\int_{V}(\vec{E}_{\text{p}}^{*}\mkern1mu{\cdot}\delta\boldsymbol\epsilon_{\text{pe}}\mkern1mu{\cdot}\vec{E}_{\text{s}})^{*}dV,\label{eq:driveneq21}
\end{equation}

where both integrals has units of $L^{-1}$.

Finally, considering the transformations: $a_{\text{p}}\rightarrow\sqrt{\hbar\omega_{\text{p}}}\,a_{\text{p}}$, $a_{\text{s}}\rightarrow\sqrt{\hbar\omega_{\text{s}}}\,a_{\text{s}}$, $b\rightarrow b_{0}\,b$ in \cref{eq:CMEopt1,eq:CMEopt2,mampeq}, assuming $\omega_{\text{s}}\approx\omega_{\text{p}}$ and imposing the Manley-Rowe conditions, we obtain the coupled mode equations in terms of the normalized amplitudes ($a_{\text{p}}$, $a_{\text{s}}$, $b$),

\begin{align}
\dot{a}_\text{p}&=i\,\Delta_{\text{p}}\,a_{\text{p}} -\,i\,g_{0}\,b\,a_{\text{s}},\label{CMEopt120}\\
\dot{a}_\text{s}&=i\,\Delta_{\text{s}}\,a_{\text{s}}-\,i\,g_{0}^{*}\,b^{*}\,a_{\text{p}},\label{CMEopt220}\\
\dot{b}&=i\,\Delta_{\text{m}}\,b-i\,g_{0}^{*}\,a_{\text{p}}\,a_{\text{s}}^{*},\label{mampeq20}
\end{align}

where, 

\begin{equation}
g_{0}=-\frac{\omega_{\text{p}}}{2}\left<\vec{u}|\vec{f}\right>^{*}x_{\text{zpf}}=g_{\text{om}}\,x_{\text{zpf}},
\label{eq:manleyR}
\end{equation}    	  
if we consider $b_{0} = 2\,x_\text{zpf}$ \cite{VanLaer:2015tz}. According to \cref{eq:driveneq21,eq:manleyR}, $g_{\text{om}}$ (and $g_{0}$) can be decomposed as a sum of two contributions, 

\begin{gather}
g_{\text{om}}^{\text{pe}}=-\frac{\omega_{\text{p}}}{2}\int_{V}\vec{E}_{\text{p}}^{*}\mkern1mu{\cdot}\delta\boldsymbol\epsilon_{\text{pe}}\mkern1mu{\cdot}\vec{E}_{\text{s}}\,dV,\label{eq:gOMPE}\\
g_{\text{om}}^{\text{mb}}=-\frac{\omega_{\text{p}}}{2}\oint_{S}[\vec{u}\mkern1mu{\cdot}\hat{n}][\delta\epsilon_{\text{mb}}[\vec{E}^{*}_{\text{p},\parallel}\mkern1mu{\cdot}\vec{E}_{\text{s},\parallel}]
-\delta\epsilon_{\text{mb}}^{-1}[\vec{D}^{*}_{\text{p},\perp}\mkern1mu{\cdot}\vec{D}_{\text{s},\perp}]]\,dA,\label{eq:gOMMB}
\end{gather}
and represent the frequency shift contributions generated by the strain per unit length, $\vec{S}$, and the moving boundary $\vec{u}\mkern1mu{\cdot}\hat{n}$ of the acoustic wave \cite{Johnson:2002tp}, respectively. The moving boundary induces permittivity fluctuations that are perceive by the optical fields. The tangential electric field is perturbed by $\delta\epsilon_{\text{mb}}$ whereas the perpendicular electric displacement field is perturbed by $\delta\epsilon_{\text{mb}}^{-1}$.\\





\section{\label{sup:fem}Finite elements method}

The eigenvalues and eigenvectors for the optical and mechanical fields are solved by using finite elements method (FEM) applied to the Helmholtz equation (\cref{eq:max2}) and the equation of motion (\cref{eq:mech_eigenmode}), respectively, both implemented in a commercial software (COMSOL 4.4). The optical modes are simulated by using the Electromagnetic Frequency Domain Interface (emw) with a modified weak form to ensure convenient field solutions whereas the mechanical modes are simulated by using the Weak Form PDE Interface (w). As the symmetry of the cavities suggest, we use cylindrical coordinates (2D-axisymmetric component in COMSOL) to simulate the structures.

\begin{figure*}[!ht]
\centering
\includegraphics[scale=1]{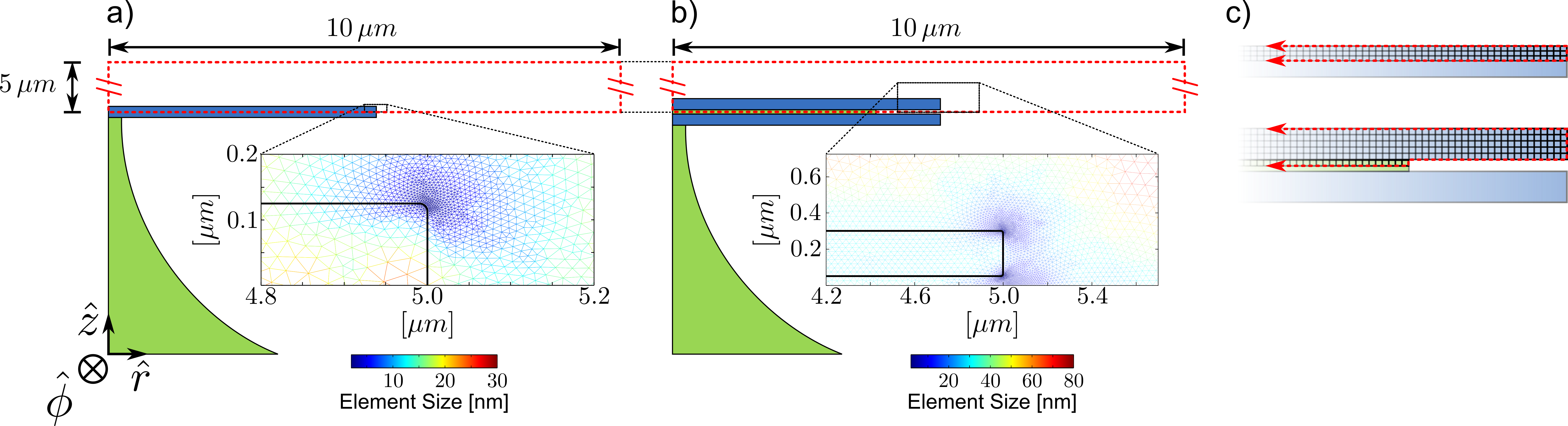}
\caption{\small{Cross section of the computational domains (internal regions defined by the red dashed lines) in the single and double disks. \textbf{a)}-\textbf{b)} Modeling of the single and double disk to calculate the coupling between the optical and mechanical modes. White, blue and green regions represent the air, silicon and silica materials, respectively. In the inset figures show the kind of mesh and the element size. In the air domain, the element size enhances radialy with a maximum growth rate of $1.1$. The top (and bottom in the double disk) right-corner is rounded, $r=10\,nm$. \textbf{c)} Modeling of the single and double disk to calculate the mechanical dispersion ($\Omega/2\pi$ vs Azimuthal wavenumber - $M$). The cartesian black grid represent the kind of mesh that is used.}}
\label{fig5}
\vskip -0.175in
\end{figure*}

We also assume a negligible effect of the pedestal on the optical and mechanical whispering gallery modes, which are mainly confined close to the circumference of the disk. This further simplify the problem with a $r-\phi$ symmetry plane, which reduces the computational domain to a half-system (half disk in the case of a simple disk and a full-disk plus half silica layer for the double-disk structure). \cref{fig5} shows the computational domains for both structures.

In order to calculate the optomechanical coupling rate the same mesh to resolve for both optical and mechanical wave equations is used. For the double disk structure the photoelastic contribution from silica to $g_{0}$ is not considered, since the slot-mode optical field is mainly confined in the air region between the silicon disks and close to the circumference. In \cref{fig5}a-b we can see the kind of mesh used in the structures. In order to to improve convergence, we employ cubic interpolation functions for optical and mechanical modes. We also use rounded disk's corners (insets in \cref{fig5}a-b), avoiding unrealistic optical fields that could impact the moving boundary overlap integrals.

In order to calculate the modal mechanical dispersion of the structures the equation of motion (\cref{eq:mech_eigenmode}) is solved by using rectangular finite elements (cartesian black grid inside of the red dashed line in \cref{fig5}c). In both structures quadratic interpolation functions are used. Matlab Livelink was used to sweep azimuthal wavenumber - $M$ parameter.

On the other hand, a perfect electric conductor boundary condition is assumed on the boundary of the computational domain to calculate the TM lowest-order mode. From mechanical point of view,  both structures are simulated like a cantilever, i. e., the left-side boundary is fixed (part of the red dashed line along to the z-axis \cref{fig5}a-b). In order to simulate dilatational and flexural modes in the single disk cavity the boundary conditions are $u_{r}\neq 0$, $u_{z}=0$, $u_{\phi}\neq 0$, and $u_{r}=0$, $u_{z}\neq 0$, $u_{\phi}=0$ in the bottom boundary (red dashed lines along to the $r$-axis \cref{fig5}c), respectively. In the double disk cavity only are applied the conditions: $u_{r}\neq 0$, $u_{z}=0$, $u_{\phi}\neq 0$.\\

\section{\label{sup:g0}Calculation of the optomechanical coupling rate: $g_0$}
In order to calculate $g_{0}$ the ansatz that is used to the \cref{eq:max2,eq:mech_eigenmode} is given by, 
\begin{align}
\vec{E}_{j}(\vec{r})&=(E_{j}^{(r)},E_{j}^{(\phi)}i,E_{j}^{(z)})\,e^{-im_{j}\phi},\label{eq:spa_opt}\\
\vec{u}(\vec{r})&=(u^{(r)},u^{(\phi)}i,u^{(z)})\,e^{-iM\phi},\label{eq:spa_mech}
\end{align}
respectively, with $j=$ p, s. We also assume that the phase-matching condition is satisfied ($M = m_{\text{p}} - m_{\text{s}}$)  and  the backscattered Stokes mode as a complementary mode~\cite{mrozowski1997guided}, i.e.,
\begin{align}
E_{\text{s}}^{(r)}&=E_{\text{p}}^{(r)}=E^{(r)},\label{eq:comp_mode0}\\
E_{\text{s}}^{(\phi)}&=-E_{\text{p}}^{(\phi)}=-E^{(\phi)},\label{eq:comp_mode1}\\
E_{\text{s}}^{(z)}&=E_{\text{p}}^{(z)}=E^{(z)},\label{eq:comp_mode}
\end{align}
The optomechanical coupling rate can be decomposed in two contributions, $g_{\text{mb}}$ and $g_{\text{pe}}$, below we detail our calculations for both contributions.
\subsection*{Moving-boundary contribution}
 For the moving boundary contribution, using  \cref{eq:spa_opt,eq:spa_mech,eq:comp_mode0,eq:comp_mode1,eq:comp_mode} and \cref{eq:gOMMB} with the relation $g_{\text{mb}}=g_{\text{om}}^{\text{mb}}x_{\text{zpf}}$, we can break up the moving boundary contribution in three terms related to each optical field component,
\begin{equation}
g_{\text{mb}}=\displaystyle\sum_{k=1}^{3}g_{\text{mb}}^{(k)},
\label{eq:gOMB_decom}
\end{equation}
where,
\begin{equation}
g_{\text{mb}}^{(k)}=-\frac{\omega_{\text{p}}\,x_{\text{zpf}}}{2}\oint_{S} u_{\perp}\rho_{\text{mb}}^{(k)}\,dA,\label{eq:g0MB_conk}
\end{equation}
with the contributions to the optical weighting function  $\rho_{\text{mb}}=\sum_{k=1}^{3}\rho_{\text{mb}}^{(k)}$ given by,
\begin{align}
\rho_{\text{mb}}^{(1)}&=\delta\epsilon_{\text{mb}}E_{\parallel}^2,\label{eq:rhomb_con0}\\
\rho_{\text{mb}}^{(2)}&=-\delta\epsilon_{\text{mb}}[E^{(\phi)}]^2,\label{eq:rhomb_con1}\\
\rho_{\text{mb}}^{(3)}&=-\delta\epsilon_{\text{mb}}^{-1}D_{\perp}^{2},\label{eq:rhomb_con}
\end{align}
the normal and tangential field and displacement components are $u_{\perp}=u^{(r)}n_{r}+u^{(z)}n_{z}$, $E_{\parallel}=E^{(r)}t_{r}+E^{(z)}t_{z}$, $D_{\perp}=D^{(r)}n_{r}+D^{(z)}n_{z}$, $\hat{n}=(n_{r},0,n_{z})$ and $\hat{t}=(t_{r},0,t_{z})$. $\hat{n}$ and $\hat{t}$ are the normal and tangential unitary vectors in the transverse $rz$-plane, respectively. The minus signal in $\rho_{\text{mb}}^{(2)}$ arise from of the Stokes $\phi$-component in \cref{eq:comp_mode1}.

Figure \ref{fig6} shows all the tangential and perpendicular electric field components and the contributions to the weighting function (\cref{eq:rhomb_con0,eq:rhomb_con1,eq:rhomb_con}) for the  lowest order TM mode in a single disk cavity. Interestingly, although $\mathit{E}_{\perp}^{2}$ is the largest optical field component, the dominant contribution to the optical weighting function is $\rho_{\text{mb}}^{(2)}$, which is proportional to the azimuthal field component. The reason why the azimuthal field dominates over the vertical field is obvious if we rewrite,
\begin{equation}
\rho_{\text{mb}}^{(3)}=\delta\epsilon_{\text{mb}}E_{\perp}^{2}\left[\frac{n_{2}}{n_{1}}\right]^{2},\label{eq:rhomb3}
\end{equation}
where $n_{1}=3.5$ and $n_{2}=1$ are the refractive indexes of the single disk cavity and the region outside of cavity, respectively. Due to the factor $(n_{2}/n_{1})^{2}\approx 0.1$ in \cref{eq:rhomb3},  the vertical component contribution is reduced by roughly one order of magnitude due to the high refractive index constrast.           

\begin{figure*}[!ht]
\centering
\includegraphics[scale=1]{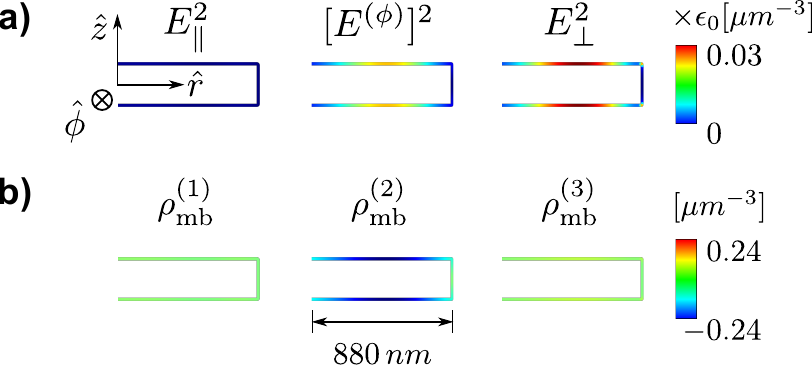}
\caption{\small{Spatial distribution of the overlapping between the TM and surface ($d_{2}$, $\Omega/2\pi=16.87\,GHz$) modes in the single disk. The mode, $d_{2}$, induces a \textbf{a)} strain ($\vec{S}$), \textbf{b)} permitivitty fluctuation ($\delta\boldsymbol\epsilon_{\text{pe}}$) and a \textbf{c)} overlapping with the TM mode ($I_{\text{pe}}^{k}$, $k=1,..,6$). As the strain is calculated from the unitary displacement $\vec{u}$, then units of the inverse to the length should be used.}}
\label{fig6}
\vskip -0.175in
\end{figure*}

In \cref{table:4} and \cref{table:3} we show each component of the moving-boundary contribution for two mechanical modes, the dilational mode $d_2$ (shown in \cref{figure4}h ) and the whispering gallery mode $w_{16}$ (show in \cref{figure4}f).\\

\begin{table}[h!]
\centering
\subfloat[$\Omega/2\pi=16.87$~GHz, $x_{\text{zpf}}=0.36$~fm, $m_\text{eff}=3.8$~pg]{
\begin{tabular}{| c | c | c | c | c |} 
 \hline
 $g_{\text{mb}}^{(1)}$ & $g_{\text{mb}}^{(2)}$ & $g_{\text{mb}}^{(3)}$ & $g_{\text{mb}}$ & $g_{\text{om}}^{\text{mb}}$\\ 
 \hline
 $+2.5$ & \textcolor{blue}{$-134$} & $+15.3$ & $-116.4$ & $-323$\\
 \hline
\end{tabular}
\label{table:4}}\hspace{2cm}
\subfloat[$\Omega/2\pi=24.34$~GHz, $x_{\text{zpf}}=0.23$~fm, $m_\text{eff}=6.3$ pg.]{
\begin{tabular}{| c | c | c | c | c |} 
 \hline
 $g_{\text{mb}}^{(1)}$ & $g_{\text{mb}}^{(2)}$ & $g_{\text{mb}}^{(3)}$ & $g_{\text{mb}}$ & $g_{\text{om}}^{\text{mb}}$\\ 
 \hline
 $-0.15$ & \textcolor{blue}{$+6.45$} & $-0.81$ & $+5.5$ & $+24$\\
 \hline
\end{tabular}
\label{table:3}
}
\caption{Moving-boundary optomechanical coupling components ($\times1/2\pi$) for the $d_{2}$ (a) and  $w_{16}$ (b) mechanical modes. Azimuthal number $M=70$, $g_{\text{om}}^{\text{mb}}$ (in GHz/nm) and $g_{\text{mb}}$ (in kHz).}
\end{table}
\subsection*{Photo-elastic contribution}
In order to grasp the nature of photoelastic component we substitute \cref{eq:spa_opt,eq:spa_mech,eq:comp_mode0,eq:comp_mode1,eq:comp_mode} in \cref{eq:gOMPE} and use the relation $g_{\text{pe}}=g_{\text{om}}^{\text{pe}}x_{\text{zpf}}$,
\begin{equation}
g_{\text{pe}}=\displaystyle\sum_{k=1}^{6}g_{\text{pe}}^{(k)},
\label{eq:gPE_contri}
\end{equation}
where,
\begin{equation}
g_{\text{pe}}^{(k)}=-\frac{\omega_{\text{p}}x_{\text{zpf}}}{2}\int_{V} I_{\text{pe}}^{(k)}dV,
\label{eq:g0PE_conk}
\end{equation}
and,
\begin{align}
I_{\text{pe}}^{(1)}&=\delta\epsilon_{\text{pe}}^{rr}[E^{(r)}]^{2},           &  I_{\text{pe}}^{(4)}&=-2iE^{(\phi)}\delta\epsilon_{\text{pe}}^{\phi z}E^{(z)},\nonumber\\
I_{\text{pe}}^{(2)}&=-\delta\epsilon_{\text{pe}}^{\phi\phi}[E^{(\phi)}]^{2},         &  I_{\text{pe}}^{(5)}&=2 E^{(r)}\delta\epsilon_{\text{pe}}^{rz}E^{(z)},\label{eq:IPE_contri}\\
I_{\text{pe}}^{(3)}&=\delta\epsilon_{\text{pe}}^{zz}[E^{(z)}]^{2},   &  I_{\text{pe}}^{(6)}&=-2iE^{(r)}\delta\epsilon_{\text{pe}}^{r \phi}E^{(\phi)},\nonumber
\end{align}
are the contributions to spatial overlap $I_{\text{pe}}=\sum_{k=1}^{6}I_{\text{pe}}^{(k)}$ and the dielectric perturbations due to photoelastic effect are given by,
\begin{align}
\delta\epsilon_{\text{pe}}^{rr}&=-\epsilon_{0}n_{1}^{4}(p_{11}S_{rr}+p_{12}\left[S_{\phi\phi}+S_{zz}\right]), & \delta\epsilon_{\text{pe}}^{\phi z}&=-\epsilon_{0}n_{1}^{4}(p_{44}S_{\phi z}),\nonumber\\
\delta\epsilon_{\text{pe}}^{\phi\phi}&=-\epsilon_{0}n_{1}^{4}(p_{11}S_{\phi\phi}+p_{12}\left[S_{rr}+S_{zz}\right]), & \delta\epsilon_{\text{pe}}^{rz}&=-\epsilon_{0}n_{1}^{4}(p_{44}S_{rz}),\label{eq:deltaPE_contri}\\
\delta\epsilon_{\text{pe}}^{zz}&=-\epsilon_{0}n_{1}^{4}(p_{11}S_{zz}+p_{12}\left[S_{rr}+S_{\phi\phi}\right]), & \delta\epsilon_{\text{pe}}^{r\phi}&=-\epsilon_{0}n_{1}^{4}(p_{44}S_{r\phi}),
\nonumber
\end{align}
where each strain tensor component is calculated as,
\begin{align}
S_{rr}&=\partial_{r}u^{(r)}, & S_{\phi z}&=\frac{i}{2}\left[\partial_{z}u^{(\phi)}-\frac{Mu^{(z)}}{r}\right],\nonumber\\
S_{\phi\phi}&=\frac{u^{(r)}+Mu^{(\phi)}}{r}, & S_{rz}&=\frac{1}{2}\left[\partial_{z}u^{(r)}+\partial_{r}u^{(z)}\right],\label{eq:Strain_con}\\
S_{zz}&=\partial_{z}u^{(z)}, & S_{r\phi}&=\frac{i}{2}\left[-\frac{Mu^{(r)}}{r}+\left[\partial_{r}-\frac{1}{r}\right]u^{(\phi)}\right].
\nonumber
\end{align}
Similarly to $g_{\text{mb}}$, $g_{\text{pe}}$ is also real.\\

\begin{figure*}[!ht]
\centering
\includegraphics[scale=1]{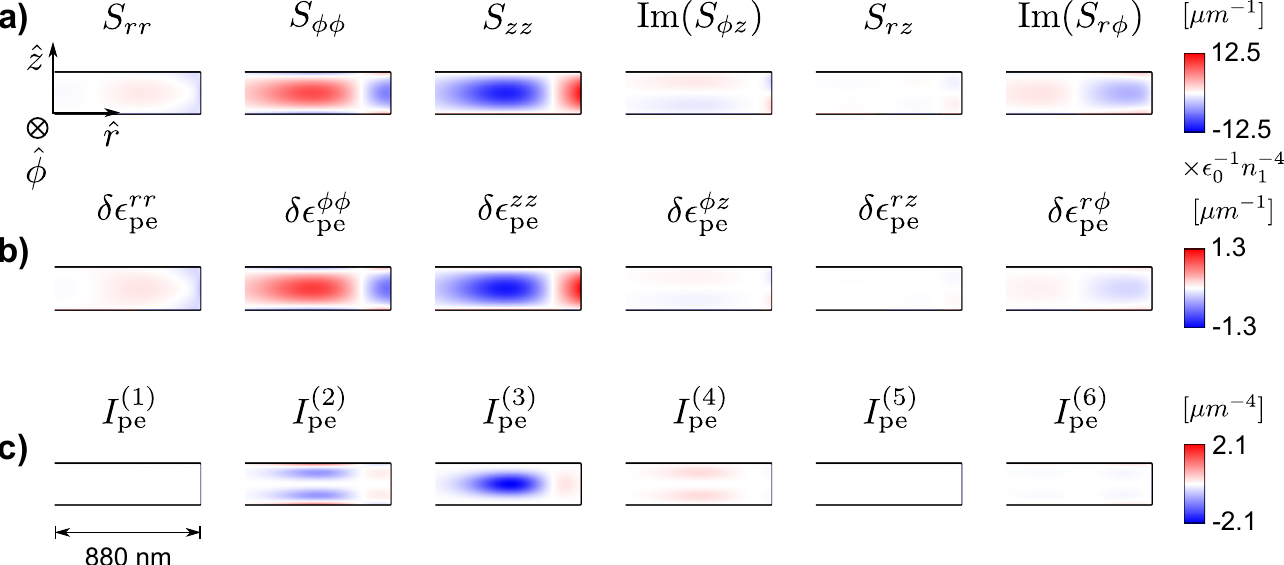}
\caption{\small{Spatial distribution of the overlapping between the TM and surface ($d_{2}$, $\Omega/2\pi=16.87\,GHz$) modes in the single disk. The mode, $d_{2}$, induces a \textbf{a)} strain ($\vec{S}$), \textbf{b)} permitivitty fluctuation ($\delta\boldsymbol\epsilon_{\text{pe}}$) and a \textbf{c)} overlap integrand ($I_{\text{pe}}^{k}$). As the strain is calculated from the unitary displacement it has units $L^{-1}$.}}
\label{fig6}
\vskip -0.175in
\end{figure*}

\begin{figure}[!ht]
\centering
\includegraphics[scale=1]{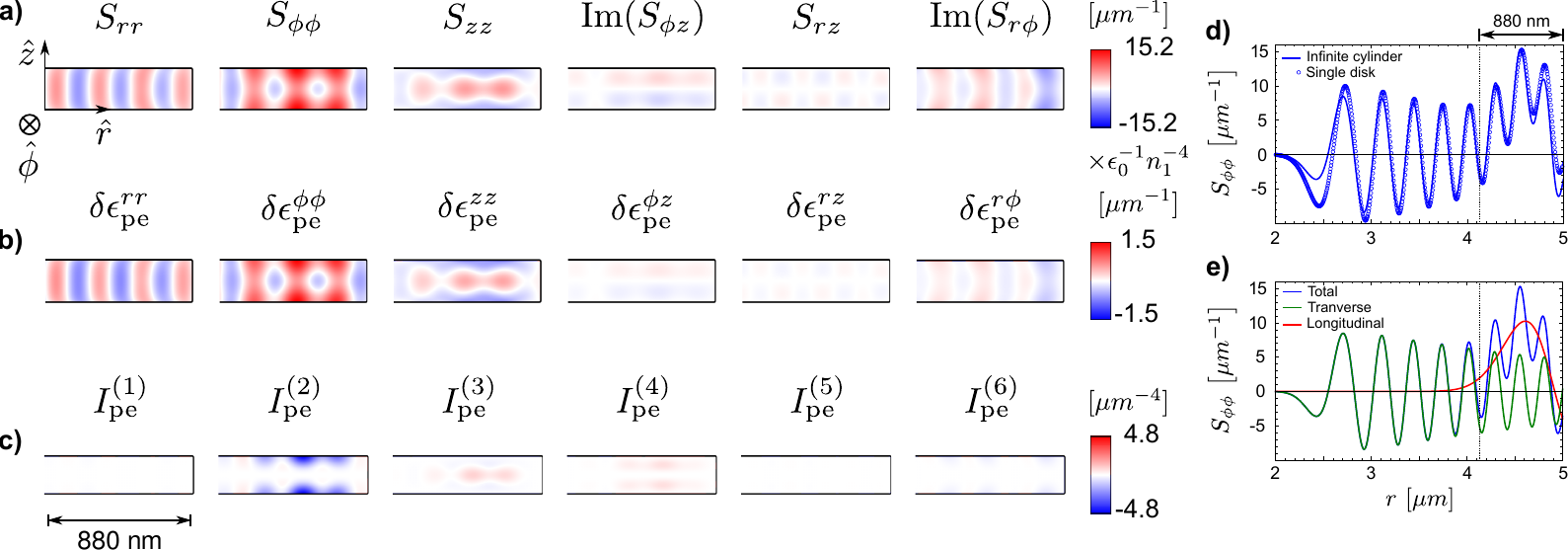}
\caption{\small{Spatial distribution of the overlap integrals between the TM and $w_{16}$ ($\Omega/2\pi=24.34\,GHz$) modes in the single disk. \textbf{a)} Mechanical strain ($\vec{S}$), \textbf{b)} Permitivitty perturbation ($\delta\boldsymbol\epsilon_{\text{pe}}$) and a \textbf{c)} overlap integrands ($I_{\text{pe}}^{k}$). \textbf{d)} Radial behavior of $S_{\phi\phi}$. Analytic (blue solid line) and simulation (blue hollow circles) data are shown to the infinite cylinder and the single disk, respectively. The single disk linecut is taken along the top plane. \textbf{e)} Analytic curves of the radial behavior of the transverse (green solid line) and longitudinal (red solid line) contributions to $S_{\phi\phi}$ (blue solid line) to the infinite cylinder.}}
\label{fig7}
\vskip -0.175in
\end{figure}

We also take the dilational mode $d_2$ (shown in \cref{figure4}h ) and the whispering gallery mode $w_{16}$ (show in \cref{figure4}f) to understand the spatial overlap behavior of \cref{eq:IPE_contri}. In Figure \ref{fig6} we show each contributions in \cref{eq:IPE_contri}, the strain components (\cref{eq:Strain_con}), and the dielectric perturbations (\cref{eq:deltaPE_contri}).  The spatial overlap $I_{\text{pe}}^{(3)}$ is dominant for the $d_{2}$ modes (\cref{fig6}), whereas the $I_{\text{pe}}^{(2)}$  is dominant for the $w_{16}$ mode (\cref{fig7}c). A similar correspondence can be observed both in the strain and dielectric perturbations.

The peaked $g_0$ contribution of the whispering mode group in the main text \cref{figure4}a is caused by the presence of a net positive azimuthal strain region close to the circumference of the single disk cavity. In  \cref{fig7}d we show this behavior in detail for the $w_{16}$ mode. The physical origin of this positive net strain region can be traced by exploring the analytical expression for $S_{\phi\phi}$ obtained for an infinite elastic cylinder.
\begin{equation}
S_{\phi\phi}=\underbrace{\frac{u^{(r)}_{\text{l}}+M u^{(\phi)}_{\text{l}}}{r}}_{S_{\phi\phi}^{\text{l}}}+\underbrace{\frac{u^{(r)}_{\text{t}}+M u^{(\phi)}_{\text{t}}}{r}}_{S_{\phi\phi}^{\text{t}}},
\label{eq:spplt}
\end{equation}
where,
\begin{align}
u^{(r)}_{\text{l}}&=-\frac{\tilde{\Omega}}{\eta }J_M'\left(\frac{\tilde{r} \tilde{\Omega} }{\eta }\right), & u^{(r)}_{\text{t}}&=\frac{M f(\tilde{\Omega}) J_M(\tilde{r} \tilde{\Omega} )}{\tilde{r}},\\
u^{(\phi)}_{\text{l}}&=\frac{M}{\tilde{r}}J_M\left(\frac{\tilde{r} \tilde{\Omega} }{\eta }\right), & u^{(\phi)}_{\text{t}}&=-\tilde{\Omega}  f(\tilde{\Omega} ) J_M'(\tilde{r} \tilde{\Omega} ),
\label{eq:urplt}
\end{align}
are the contributions from the longitudinal (l) and transverse (t) waves to each displacement component \cite{Dmitriev2014905},
\begin{equation}
f(\tilde{\Omega})=\frac{1}{\eta ^2}\frac{J_{M-2}\left(\frac{\tilde{\Omega} }{\eta }\right)-J_{M+2}\left(\frac{\tilde{\Omega} }{\eta }\right)}{J_{M-2}(\tilde{\Omega} )+J_{M+2}(\tilde{\Omega} )},
\label{eq:fomega}
\end{equation}
where $J_{M}$ is the Bessel function of the first kind of order $M$, $\tilde{\Omega}=\frac{\Omega_{0}^{\text{c}}a}{V_{\text{t}}}$ is the normalized angular frequency; $\Omega_{0}^{\text{c}}$ is the angular frequency, $a$ is the cylinder radius and the transverse bulk velocity is $V_{\text{t}}$, $\eta=\frac{V_{\text{l}}}{V_{\text{t}}}$; $V_{\text{l}}$ is the longitudinal bulk velocity and $\tilde{r}=r/a$ is the normalized radius.

There is a surprisingly good agreement between the analytic  (blue solid line) mode profile and the actual numerical mode for the microdisk (blue hollow circles) in the \cref{fig7}d. The analytical solution has an explicit contribution from the longitudinal and tranverse propagation velocities. The slowly varying contribution, is due to the slower radial wavevector associated with the longitudinal wave. Indeed with we plot just this contribution in the analytical solution we can precisely reproduce the bump observed in the numerical solution(\cref{fig7}e). Therefore we attribute the slowly varying positive net strain to the contrasting velocities of transverse and longitudinal acoustic waves in Si.\\  

In \cref{table:1} and \cref{table:2} we also show each component of the photo-elastic contribution for the two mechanical modes discussed in \cref{fig6} and \cref{fig7}, the dilational mode $d_2$ and the whispering gallery mode $w_{16}$. In both tables the dominant contributions (values in blue color) reflect the overlaps functions, as expected. We see that $g_{\text{pe}}(d_{2})$ is 67\% greater than $g_{\text{pe}}(w_{16})$, which it is not true for the dominant contributions $g_{\text{pe}}^{(3)}(d_{2})$ and $g_{\text{pe}}^{(2)}(w_{16})$.\\
\begin{table}[h!]
\centering
\subfloat[$\Omega/2\pi=16.87$~GHz, $x_{\text{zpf}}=0.36$~fm, $m_{eff}=3.8$~pg]{
\begin{tabular}{| c | c | c | c | c | c | c | c |} 
 \hline
 $g_{\text{pe}}^{(1)}$ & $g_{\text{pe}}^{(2)}$ & $g_{\text{pe}}^{(3)}$ & $g_{\text{pe}}^{(4)}$ & $g_{\text{pe}}^{(5)}$ & $g_{\text{pe}}^{(6)}$ & $g_{\text{pe}}$ & $g_{\text{om}}^{\text{pe}}$\\ 
 \hline
 $+0.02$ & $+22.2$ & \textcolor{blue}{$+78.8$} & $-14.7$ & $-0.3$ & $+1.0$ & $+87.2$ & $+242$\\
 \hline
\end{tabular}
\label{table:2}}\hspace{2cm}
\subfloat[$\Omega/2\pi=24.34$~GHz, $x_{\text{zpf}}=0.23$~fm, $m_{eff}=6.3$~pg]{
\begin{tabular}{| c | c | c | c | c | c | c | c |} 
 \hline
 $g_{\text{pe}}^{(1)}$ & $g_{\text{pe}}^{(2)}$ & $g_{\text{pe}}^{(3)}$ & $g_{\text{pe}}^{(4)}$ & $g_{\text{pe}}^{(5)}$ & $g_{\text{pe}}^{(6)}$ & $g_{\text{pe}}$ & $g_{\text{om}}^{\text{pe}}$\\ 
 \hline
 $+0.2$ & \textcolor{blue}{$+79.3$} & $-7.9$ & $-16.6$ & $-0.4$ & $+6.6$ & $+61.2$ & $+266$\\
 \hline
\end{tabular}
\label{table:1}
}
\caption{Photo-elastic optomechanical coupling components ($\times1/2\pi$) for the $d_{2}$ (a) and  $w_{16}$ (b) mechanical modes. Azimuthal number $M=70$, $g_{\text{om}}^{\text{pe}}$ (in GHz/nm) and $g_{\text{pe}}$ (in kHz).}
\end{table}
\section{\label{sup:thresh}Brillouin lasing threshold}
In order to calculate the power threshold we take the \cref{CMEopt120,CMEopt220,mampeq20} and add the losses ($\kappa_{\text{e}}$, $\kappa_{\text{p}}$, $\kappa_{\text{s}}$, $\Gamma$), the normalized power amplitude ($s_{\text{p}}\rightarrow\frac{s_{\text{p}}}{\sqrt{\hbar\omega_{\text{p}}}}$) and considering that $g_{0}\rightarrow g_{0}^{\text{c}}=(g_{0}^{\text{c}})^{*}$, where $g_{0}^{\text{c}}$ is the vacuum optomechanical coupling rate for the compound cavity,

\begin{align}
\dot{a}_\text{p}&=\chi_\text{p}^{-1} a_\text{p} -i\,g_{0}^{\text{c}}\,b\,a_\text{s}+\sqrt{\kappa_\text{e}}s_{\text{p}},\label{eq:cme_tP}\\
\dot{a}_\text{s}&=\chi_\text{s}^{-1} a_\text{s} -i\,g_{0}^{\text{c}}\,b^{*} a_\text{p},\label{eq:cme_tS}\\
\dot{b}&=\chi_\text{m}^{-1}b-i\,g_{0}^{\text{c}}\,a_{\text{p}}a_{\text{s}}^{*},
\label{eq:cme_tb}
\end{align}


in which $\chi_\text{p}^{-1}=i\Delta_{\text{p}}+\frac{\kappa_{\text{p}}}{2}$, $\chi_\text{s}^{-1}=i\Delta_{\text{s}}+\frac{\kappa_{\text{s}}}{2}$ and $\chi_\text{m}^{-1}=i\Delta_{\text{m}}+\frac{\Gamma}{2}$. Now following~\cite{Matsko:2002vs}, the steady-state in the \cref{eq:cme_tP,eq:cme_tb} leads to,

\begin{equation}
\dot{a}_\text{s}=\left[\chi_{\text{s}}^{-1}-\frac{\kappa_{\text{e}}|s_{\text{p}}|^{2}(g_{0}^{\text{c}})^{2}}{[\chi_{\text{m}}^{-1}]^{*}|\chi_{\text{p}}^{-1}|^{2}}\left|1+\frac{|a_{\text{s}}|^{2}(g_{0}^{\text{c}})^{2}}{\chi_{\text{m}}^{-1}\chi_{\text{p}}^{-1}}\right|^{-2}\right]a_{\text{s}},
\label{eq:ss1}
\end{equation}

in which to reach non-trivial steady-state the term between parentheses should be zero and as a consequence results the threshold condition,

\begin{equation}
|s_{\text{p}}|^{2}>\frac{[\chi_{\text{m}}^{-1}]^{*}\chi_{\text{s}}^{-1}|\chi_{\text{p}}^{-1}|^{2}}{\kappa_{\text{e}}(g_{0}^{\text{c}})^{2}}.
\label{eq:threhold_p}
\end{equation}

From the \cref{eq:threhold_p} we have a product between two complex variables: $[\chi_{\text{m}}^{-1}]^{*}\chi_{\text{s}}^{-1}$. In order to understand the nature of this product we come back to the expression between parentheses in the \cref{eq:ss1} in the steady-state and rewrite,

\begin{equation}
\chi_{\text{s}}^{-1}=\frac{\kappa_{\text{e}}|s_{\text{p}}|^{2}(g_{0}^{\text{c}})^{2}}{|\chi_{\text{m}}^{-1}|^{2}|\chi_{\text{p}}^{-1}|^{2}}\left|1+\frac{|a_{\text{s}}|^{2}(g_{0}^{\text{c}})^{2}}{\chi_{\text{m}}^{-1}\chi_{\text{p}}^{-1}}\right|^{-2}\chi_{\text{m}}^{-1},
\label{eq:ss}
\end{equation}

in which substituting the expressions to $\chi_{\text{s}}^{-1}$ and $\chi_{\text{m}}^{-1}$ and simplifying, results,

\begin{equation}
\frac{\Delta_{\text{s}}}{\kappa_{\text{s}}}=\frac{\Delta_{\text{m}}}{\Gamma}.
\label{eq:kappa_Gamma}
\end{equation}

Now by using the \cref{eq:kappa_Gamma} and the expression to $\chi_{\text{p}}^{-1}$ in the threshold condition (\cref{eq:threhold_p}) we obtain $|s_{\text{p}}|^{2}>P_\text{th}$,

\begin{equation}
P_\text{th}=\frac{\hbar\omega_{\text{p}}\kappa_{\text{p}}^{2}}{4\,\mathcal{C}\kappa_{\text{e}}}\left[1+\left(\frac{\Delta_{\text{s}}}{\kappa_{\text{s}}/2}\right)^{2}\right]\left[1+\left(\frac{\Delta_{\text{p}}}{\kappa_{\text{p}}/2}\right)^{2}\right],
\label{eq:threhold_p2}
\end{equation}

in which $\mathcal{C}=\frac{4(g_{0}^{\text{c}})^{2}}{\Gamma\kappa_{\text{s}}}$ is the so-called single-photon cooperativity.


\begin{thebibliography}{10}
\expandafter\ifx\csname url\endcsname\relax
  \def\url#1{\texttt{#1}}\fi
\expandafter\ifx\csname urlprefix\endcsname\relax\def\urlprefix{URL }\fi
\providecommand{\bibinfo}[2]{#2}
\providecommand{\eprint}[2][]{\url{#2}}

\bibitem{Boyd:2050257}
\bibinfo{author}{Boyd, R.~W.}
\newblock \emph{\bibinfo{title}{{Nonlinear optics}}}
  (\bibinfo{publisher}{Elsevier Science}, \bibinfo{address}{Burlington, MA},
  \bibinfo{year}{2013}).

\bibitem{Johnson:2002tp}
\bibinfo{author}{Johnson, S.} \emph{et~al.}
\newblock \bibinfo{title}{{Perturbation theory for Maxwell{\textquoteright}s
  equations with shifting material boundaries}}.
\newblock \emph{\bibinfo{journal}{Physical Review E}}
  \textbf{\bibinfo{volume}{65}}, \bibinfo{pages}{066611}
  (\bibinfo{year}{2002}).

\bibitem{Chan:2011dy}
\bibinfo{author}{Chan, J.} \emph{et~al.}
\newblock \bibinfo{title}{{Laser cooling of a nanomechanical oscillator into
  its quantum ground state}}.
\newblock \emph{\bibinfo{journal}{Nature}} \textbf{\bibinfo{volume}{478}},
  \bibinfo{pages}{89--92} (\bibinfo{year}{2011}).

\bibitem{Anonymous:ARK9m6Hv}
\bibinfo{author}{Safavi-Naeini, A.~H.} \emph{et~al.}
\newblock \bibinfo{title}{{Observation of Quantum Motion of a Nanomechanical
  Resonator}}.
\newblock \emph{\bibinfo{journal}{Physical Review Letters}}
  \textbf{\bibinfo{volume}{108}}, \bibinfo{pages}{033602}
  (\bibinfo{year}{2012}).

\bibitem{Li:2012bfa}
\bibinfo{author}{Li, J.}, \bibinfo{author}{Lee, H.}, \bibinfo{author}{Chen, T.}
  \& \bibinfo{author}{Vahala, K.~J.}
\newblock \bibinfo{title}{{Characterization of a high coherence, Brillouin
  microcavity laser on silicon}}.
\newblock \emph{\bibinfo{journal}{Optics Express}}
  \textbf{\bibinfo{volume}{20}}, \bibinfo{pages}{20170--20180}
  (\bibinfo{year}{2012}).

\bibitem{Smith:1991co}
\bibinfo{author}{Smith, S.~P.}, \bibinfo{author}{Zarinetchi, F.} \&
  \bibinfo{author}{Ezekiel, S.}
\newblock \bibinfo{title}{{Narrow-linewidth stimulated Brillouin fiber laser
  and applications}}.
\newblock \emph{\bibinfo{journal}{Optics Letters}}
  \textbf{\bibinfo{volume}{16}}, \bibinfo{pages}{393--395}
  (\bibinfo{year}{1991}).

\bibitem{Debut:2000jz}
\bibinfo{author}{Debut, A.}, \bibinfo{author}{Randoux, S.} \&
  \bibinfo{author}{Zemmouri, J.}
\newblock \bibinfo{title}{{Linewidth narrowing in Brillouin lasers: Theoretical
  analysis}}.
\newblock \emph{\bibinfo{journal}{Physical Review A}}
  \textbf{\bibinfo{volume}{62}}, \bibinfo{pages}{023803}
  (\bibinfo{year}{2000}).

\bibitem{Gross:2010jg}
\bibinfo{author}{Gross, M.~C.} \emph{et~al.}
\newblock \bibinfo{title}{{Tunable millimeter-wave frequency synthesis up to
  100 GHz by dual-wavelength Brillouin fiber laser}}.
\newblock \emph{\bibinfo{journal}{Optics Express}}
  \textbf{\bibinfo{volume}{18}}, \bibinfo{pages}{13321--13330}
  (\bibinfo{year}{2010}).

\bibitem{Marpaung:2014vf}
\bibinfo{author}{Marpaung, D.}, \bibinfo{author}{Pagani, M.},
  \bibinfo{author}{Morrison, B.} \& \bibinfo{author}{Eggleton, B.~J.}
\newblock \bibinfo{title}{{Nonlinear Integrated Microwave Photonics}}.
\newblock \emph{\bibinfo{journal}{Journal of Lightwave Technology}}
  \textbf{\bibinfo{volume}{32}}, \bibinfo{pages}{3421--3427}
  (\bibinfo{year}{2014}).

\bibitem{Dainese:2006ta}
\bibinfo{author}{Dainese, P.} \emph{et~al.}
\newblock \bibinfo{title}{{Raman-like light scattering from acoustic phonons in
  photonic crystal fiber}}.
\newblock \emph{\bibinfo{journal}{Optics Express}}
  \textbf{\bibinfo{volume}{14}}, \bibinfo{pages}{4141--4150}
  (\bibinfo{year}{2006}).

\bibitem{Dainese:2006tj}
\bibinfo{author}{Dainese, P.} \emph{et~al.}
\newblock \bibinfo{title}{{Stimulated Brillouin scattering from
  multi-GHz-guided acoustic phonons in nanostructured photonic crystal
  fibres}}.
\newblock \emph{\bibinfo{journal}{Nature Physics}}
  \textbf{\bibinfo{volume}{2}}, \bibinfo{pages}{388} (\bibinfo{year}{2006}).

\bibitem{Wiederhecker:2008tu}
\bibinfo{author}{Wiederhecker, G.~S.}, \bibinfo{author}{Brenn, A.},
  \bibinfo{author}{Fragnito, H.} \& \bibinfo{author}{Russell, P.}
\newblock \bibinfo{title}{{Coherent control of ultrahigh-frequency acoustic
  resonances in photonic crystal fibers}}.
\newblock \emph{\bibinfo{journal}{Physical Review Letters}}
  \textbf{\bibinfo{volume}{100}}, \bibinfo{pages}{203903}
  (\bibinfo{year}{2008}).

\bibitem{Kang:2009dja}
\bibinfo{author}{Kang, M.~S.}, \bibinfo{author}{Nazarkin, A.},
  \bibinfo{author}{Brenn, A.} \& \bibinfo{author}{Russell, P. S.~J.}
\newblock \bibinfo{title}{{Tightly trapped acoustic phonons in photonic crystal
  fibres as highly nonlinear artificial Raman|[nbsp]|oscillators}}.
\newblock \emph{\bibinfo{journal}{Nature Physics}}
  \textbf{\bibinfo{volume}{5}}, \bibinfo{pages}{276--280}
  (\bibinfo{year}{2009}).

\bibitem{Pant:2011ih}
\bibinfo{author}{Pant, R.} \emph{et~al.}
\newblock \bibinfo{title}{{On-chip stimulated Brillouin scattering}}.
\newblock \emph{\bibinfo{journal}{Optics Express}}
  \textbf{\bibinfo{volume}{19}}, \bibinfo{pages}{8285} (\bibinfo{year}{2011}).

\bibitem{Rakich:2012et}
\bibinfo{author}{Rakich, P.~T.}, \bibinfo{author}{Reinke, C.},
  \bibinfo{author}{Camacho, R.}, \bibinfo{author}{Davids, P.} \&
  \bibinfo{author}{Wang, Z.}
\newblock \bibinfo{title}{{Giant enhancement of stimulated Brillouin scattering
  in the subwavelength limit}}.
\newblock \emph{\bibinfo{journal}{Physical Review X}}
  \textbf{\bibinfo{volume}{2}}, \bibinfo{pages}{011008} (\bibinfo{year}{2012}).

\bibitem{Shin:2013fr}
\bibinfo{author}{Shin, H.} \emph{et~al.}
\newblock \bibinfo{title}{{Tailorable stimulated Brillouin scattering in
  nanoscale silicon waveguides}}.
\newblock \emph{\bibinfo{journal}{Nature Communications}}
  \textbf{\bibinfo{volume}{4}} (\bibinfo{year}{2013}).

\bibitem{VanLaer:2015jk}
\bibinfo{author}{Van~Laer, R.}, \bibinfo{author}{Kuyken, B.},
  \bibinfo{author}{Van~Thourhout, D.} \& \bibinfo{author}{Baets, R.}
\newblock \bibinfo{title}{{Interaction between light and highly confined
  hypersound in a silicon photonic nanowire}}.
\newblock \emph{\bibinfo{journal}{Nature Photonics}}
  \textbf{\bibinfo{volume}{9}}, \bibinfo{pages}{199--203}
  (\bibinfo{year}{2015}).

\bibitem{Wolff:2015ji}
\bibinfo{author}{Wolff, C.}, \bibinfo{author}{Steel, M.~J.},
  \bibinfo{author}{Eggleton, B.~J.} \& \bibinfo{author}{Poulton, C.~G.}
\newblock \bibinfo{title}{{Stimulated Brillouin scattering in integrated
  photonic waveguides: Forces, scattering mechanisms, and coupled-mode
  analysis}}.
\newblock \emph{\bibinfo{journal}{Physical Review A}}
  \textbf{\bibinfo{volume}{92}} (\bibinfo{year}{2015}).

\bibitem{Lee:2012hn}
\bibinfo{author}{Lee, H.} \emph{et~al.}
\newblock \bibinfo{title}{{Chemically etched ultrahigh-Q wedge-resonator on a
  silicon chip}}.
\newblock \emph{\bibinfo{journal}{Nature Photonics}}
  \textbf{\bibinfo{volume}{6}}, \bibinfo{pages}{369--373}
  (\bibinfo{year}{2012}).

\bibitem{Grudinin:2009io}
\bibinfo{author}{Grudinin, I.~S.}, \bibinfo{author}{Matsko, A.~B.} \&
  \bibinfo{author}{Maleki, L.}
\newblock \bibinfo{title}{{Brillouin Lasing with a CaF2Whispering Gallery Mode
  Resonator}}.
\newblock \emph{\bibinfo{journal}{Physical Review Letters}}
  \textbf{\bibinfo{volume}{102}}, \bibinfo{pages}{043902}
  (\bibinfo{year}{2009}).

\bibitem{Lin:2014kc}
\bibinfo{author}{Lin, G.} \emph{et~al.}
\newblock \bibinfo{title}{{Cascaded Brillouin lasing in monolithic barium
  fluoride whispering gallery mode resonators}}.
\newblock \emph{\bibinfo{journal}{Applied Physics Letters}}
  \textbf{\bibinfo{volume}{105}}, \bibinfo{pages}{231103}
  (\bibinfo{year}{2014}).

\bibitem{Bahl:2012hf}
\bibinfo{author}{Bahl, G.}, \bibinfo{author}{Fan, X.} \&
  \bibinfo{author}{Carmon, T.}
\newblock \bibinfo{title}{{Acoustic whispering-gallery modes in optomechanical
  shells}}.
\newblock \emph{\bibinfo{journal}{New Journal of Physics}}
  \textbf{\bibinfo{volume}{14}}, \bibinfo{pages}{115026}
  (\bibinfo{year}{2012}).

\bibitem{Bahl:2013eb}
\bibinfo{author}{Bahl, G.} \emph{et~al.}
\newblock \bibinfo{title}{{Brillouin cavity optomechanics with microfluidic
  devices}}.
\newblock \emph{\bibinfo{journal}{Nature Communications}}
  \textbf{\bibinfo{volume}{4}} (\bibinfo{year}{2013}).

\bibitem{Bahl:2012jm}
\bibinfo{author}{Bahl, G.}, \bibinfo{author}{Tomes, M.},
  \bibinfo{author}{Marquardt, F.} \& \bibinfo{author}{Carmon, T.}
\newblock \bibinfo{title}{{Observation of spontaneous Brillouin cooling}}.
\newblock \emph{\bibinfo{journal}{Nature Physics}}
  \textbf{\bibinfo{volume}{8}}, \bibinfo{pages}{203--207}
  (\bibinfo{year}{2012}).

\bibitem{Tomes:2009iy}
\bibinfo{author}{Tomes, M.} \& \bibinfo{author}{Carmon, T.}
\newblock \bibinfo{title}{{Photonic Micro-Electromechanical Systems Vibrating
  at X-band (11-GHz) Rates}}.
\newblock \emph{\bibinfo{journal}{Phys. Rev. Lett.}}
  \textbf{\bibinfo{volume}{102}}, \bibinfo{pages}{113601}
  (\bibinfo{year}{2009}).

\bibitem{Florez:2016aa}
\bibinfo{author}{Florez, O.} \emph{et~al.}
\newblock \bibinfo{title}{Brillouin scattering self-cancellation}.
\newblock \emph{\bibinfo{journal}{Nature Communications}}
  \textbf{\bibinfo{volume}{7}}, \bibinfo{pages}{11759 EP --}
  (\bibinfo{year}{2016}).
\newblock \urlprefix\url{http://dx.doi.org/10.1038/ncomms11759}.

\bibitem{Tamura:2009bx}
\bibinfo{author}{Tamura, S.-i.}
\newblock \bibinfo{title}{{Vibrational cavity modes in a free cylindrical
  disk}}.
\newblock \emph{\bibinfo{journal}{Physical Review B}}
  \textbf{\bibinfo{volume}{79}}, \bibinfo{pages}{054302}
  (\bibinfo{year}{2009}).
\newblock \urlprefix\url{http://link.aps.org/doi/10.1103/PhysRevB.79.054302}.

\bibitem{Sturman:2015jy}
\bibinfo{author}{Sturman, B.} \& \bibinfo{author}{Breunig, I.}
\newblock \bibinfo{title}{{Acoustic whispering gallery modes within the theory
  of elasticity}}.
\newblock \emph{\bibinfo{journal}{Journal of Applied Physics}}
  \textbf{\bibinfo{volume}{118}}, \bibinfo{pages}{013102}
  (\bibinfo{year}{2015}).

\bibitem{Matsko:2005cz}
\bibinfo{author}{Matsko, A.~B.}, \bibinfo{author}{Savchenkov, A.~A.},
  \bibinfo{author}{Strekalov, D.} \& \bibinfo{author}{Maleki, L.}
\newblock \bibinfo{title}{{Whispering Gallery Resonators for Studying Orbital
  Angular Momentum of a Photon}}.
\newblock \emph{\bibinfo{journal}{Physical Review Letters}}
  \textbf{\bibinfo{volume}{95}}, \bibinfo{pages}{143904}
  (\bibinfo{year}{2005}).

\bibitem{Dostart:2015ju}
\bibinfo{author}{Dostart, N.}, \bibinfo{author}{Kim, S.} \&
  \bibinfo{author}{Bahl, G.}
\newblock \bibinfo{title}{{Giant gain enhancement in surface-confined resonant
  Stimulated Brillouin Scattering}}.
\newblock \emph{\bibinfo{journal}{Laser {\&} Photonics Reviews}}
  \bibinfo{pages}{n/a--n/a} (\bibinfo{year}{2015}).

\bibitem{Matsko:2002vs}
\bibinfo{author}{Matsko, A.~B.}, \bibinfo{author}{Ilchenko, V.~S.},
  \bibinfo{author}{Savchenkov, A.~A.} \& \bibinfo{author}{Maleki, L.}
\newblock \bibinfo{title}{{Highly nondegenerate all-resonant optical parametric
  oscillator}}.
\newblock \emph{\bibinfo{journal}{Physical Review A}}
  \textbf{\bibinfo{volume}{66}}, \bibinfo{pages}{043814}
  (\bibinfo{year}{2002}).

\bibitem{Agarwal:2013hl}
\bibinfo{author}{Agarwal, G.~S.} \& \bibinfo{author}{Jha, S.~S.}
\newblock \bibinfo{title}{{Multimode phonon cooling via three-wave parametric
  interactions with optical fields}}.
\newblock \emph{\bibinfo{journal}{Physical Review A}}  (\bibinfo{year}{2013}).

\bibitem{Grudinin:2009tm}
\bibinfo{author}{Grudinin, I.~S.}, \bibinfo{author}{Lee, H.},
  \bibinfo{author}{Painter, O.~J.} \& \bibinfo{author}{Vahala, K.~J.}
\newblock \bibinfo{title}{{Phonon Laser Action in a Tunable Two-Level System}}.
\newblock \emph{\bibinfo{journal}{Physical Review Letters}}
  \textbf{\bibinfo{volume}{104}} (\bibinfo{year}{2010}).

\bibitem{Dmitriev2014905}
\bibinfo{author}{Dmitriev, A.}, \bibinfo{author}{Gritsenko, D.} \&
  \bibinfo{author}{Mitrofanov, V.}
\newblock \bibinfo{title}{Surface vibrational modes in disk-shaped resonators}.
\newblock \emph{\bibinfo{journal}{Ultrasonics}} \textbf{\bibinfo{volume}{54}},
  \bibinfo{pages}{905 -- 913} (\bibinfo{year}{2014}).
\newblock
  \urlprefix\url{http://www.sciencedirect.com/science/article/pii/S0041624X13003260}.

\bibitem{Wiederhecker:2009ex}
\bibinfo{author}{Wiederhecker, G.~S.}, \bibinfo{author}{Chen, L.},
  \bibinfo{author}{Gondarenko, A.} \& \bibinfo{author}{Lipson, M.}
\newblock \bibinfo{title}{{Controlling photonic structures using optical
  forces}}.
\newblock \emph{\bibinfo{journal}{Nature}} \textbf{\bibinfo{volume}{462}},
  \bibinfo{pages}{633--U103} (\bibinfo{year}{2009}).

\bibitem{Zhang:2012us}
\bibinfo{author}{Zhang, M.} \emph{et~al.}
\newblock \bibinfo{title}{{Synchronization of Micromechanical Oscillators Using
  Light}}.
\newblock \emph{\bibinfo{journal}{Phys. Rev. Lett.}}
  \textbf{\bibinfo{volume}{109}}, \bibinfo{pages}{233906}
  (\bibinfo{year}{2012}).

\bibitem{Borselli:2004ds}
\bibinfo{author}{Borselli, M.}, \bibinfo{author}{Srinivasan, K.},
  \bibinfo{author}{Barclay, P.~E.} \& \bibinfo{author}{Painter, O.~J.}
\newblock \bibinfo{title}{{Rayleigh scattering, mode coupling, and optical loss
  in silicon microdisks}}.
\newblock \emph{\bibinfo{journal}{Applied Physics Letters}}
  \textbf{\bibinfo{volume}{85}}, \bibinfo{pages}{3693} (\bibinfo{year}{2004}).
\newblock
  \urlprefix\url{http://scitation.aip.org/content/aip/journal/apl/85/17/10.1063/1.1811378}.

\bibitem{Haus:1991aa}
\bibinfo{author}{Haus, H.} \& \bibinfo{author}{Huang, W.}
\newblock \bibinfo{title}{Coupled-mode theory}.
\newblock \emph{\bibinfo{journal}{Proceedings of the IEEE}}
  \textbf{\bibinfo{volume}{79}}, \bibinfo{pages}{1505--1518}
  (\bibinfo{year}{1991}).

\bibitem{hopcroft2010young}
\bibinfo{author}{Hopcroft, M.}, \bibinfo{author}{Nix, W.~D.},
  \bibinfo{author}{Kenny, T.~W.} \emph{et~al.}
\newblock \bibinfo{title}{What is the young's modulus of silicon?}
\newblock \emph{\bibinfo{journal}{Microelectromechanical Systems, Journal of}}
  \textbf{\bibinfo{volume}{19}}, \bibinfo{pages}{229--238}
  (\bibinfo{year}{2010}).

\bibitem{Biegelsen:1974}
\bibinfo{author}{Biegelsen, D.~K.}
\newblock \bibinfo{title}{Photoelastic tensor of silicon and the volume
  dependence of the average gap.}
\newblock \emph{\bibinfo{journal}{Phys. Rev. Lett.}}
  \textbf{\bibinfo{volume}{33}}, \bibinfo{pages}{51--51}
  (\bibinfo{year}{1974}).
\newblock \urlprefix\url{http://link.aps.org/doi/10.1103/PhysRevLett.33.51}.

\bibitem{panofsky2012classical}
\bibinfo{author}{Panofsky, W.} \& \bibinfo{author}{Phillips, M.}
\newblock \emph{\bibinfo{title}{Classical Electricity and Magnetism: Second
  Edition}}.
\newblock Dover Books on Physics (\bibinfo{publisher}{Dover Publications},
  \bibinfo{year}{2012}).
\newblock \urlprefix\url{https://books.google.com.br/books?id=izEx8V4GZoEC}.

\bibitem{VanLaer:2015tz}
\bibinfo{author}{Van~Laer, R.}, \bibinfo{author}{Baets, R.} \&
  \bibinfo{author}{Van~Thourhout, D.}
\newblock \bibinfo{title}{{Unifying Brillouin scattering and cavity
  optomechanics}}  (\bibinfo{year}{2015}).
\newblock \eprint{1503.03044}.

\bibitem{mrozowski1997guided}
\bibinfo{author}{Mrozowski, M.}
\newblock \emph{\bibinfo{title}{Guided Electromagnetic Waves: Properties and
  Analysis}}.
\newblock Computer methods in electromagnetics series
  (\bibinfo{publisher}{Research Studies Press}, \bibinfo{year}{1997}).
\newblock \urlprefix\url{https://books.google.com.br/books?id=EHltQgAACAAJ}.

\end{thebibliography}
\end{document}